\documentclass[longauth]{aa}
\usepackage{placeins}
\usepackage{graphicx}
\usepackage{numprint}
\usepackage{txfonts}
\usepackage[colorlinks=true,
            linkcolor=black,
            urlcolor=blue,
            citecolor=blue]{hyperref}
\usepackage{xspace}
\usepackage{ulem}

\newcommand{\system}{WASP-33\xspace}
\newcommand{\planet}{WASP-33\,b\xspace}

\newcommand{\feh}{\mbox{[Fe/H]}\xspace}
\newcommand{\teff}{\mbox{$T_{\rm \star, eff}$}\xspace}
\newcommand{\logg}{\mbox{$\log g_\star$}\xspace}

\newcommand{\vsini}{\mbox{$v \sin i_{\star}$}\xspace}
\newcommand{\kms}{\mbox{km\,s$^{-1}$}\xspace}

\newcommand{\rplanet}{\mbox{$R_{\rm p}$}\xspace}
\newcommand{\mjup}{\mbox{$\mathrm{M_{\rm Jup}}$}\xspace}
\newcommand{\rjup}{\mbox{$\mathrm{R_{\rm Jup}}$}\xspace}

\newcommand{\mstar}{\mbox{$M_{\star}$}\xspace}
\newcommand{\rstar}{\mbox{$R_{\star}$}\xspace}

\newcommand{\msol}{\mbox{$\mathrm{M_\odot}$}\xspace}
\newcommand{\rsol}{\mbox{$\mathrm{R_\odot}$}\xspace}

\newcommand{\prot}{\mbox{$P_\mathrm{\star,rot}$}\xspace}

\newcommand{\tlcm}{{\sc TLCM}\xspace}

\newcommand{\iplanet}{\mbox{$i_{\rm p}$}\xspace}
\newcommand{\istar}{\mbox{$i_{\star}$}\xspace}
\newcommand{\Iplanet}{\mbox{$I_{\rm p}$}\xspace}
\newcommand{\Istar}{\mbox{$I_{\star}$}\xspace}
\newcommand{\porb}{\mbox{$P_\mathrm{orb}$}\xspace}
\newcommand{\Omstar}{\mbox{$\Omega_{\star}$}\xspace}
\newcommand{\Omplanet}{\mbox{$\Omega_{\rm p}$}\xspace}
\newcommand{\ktwostar}{\mbox{$k_\mathrm{2,\star}$}\xspace}
\newcommand{\agestar}{\mbox{$\tau_{\star}$}\xspace}

\newcommand{\PERIODA}{$1.2198747 \pm 0.0000012 $}
\newcommand{\EPOCHA}{$8792.63255 \pm 0.00053$}
\newcommand{\ARSTARA}{$3.6066282279 \substack{+0.036 \\ -0.045}$}
\newcommand{\RPRSA}{$0.10918 \pm 0.00098$}
\newcommand{\IMPACTA}{$0.10\substack{+0.10 \\ -0.08}$}
\newcommand{\UPLUSA}{$0.801\substack{+0.087 \\ -0.090}$}
\newcommand{\UMINUSA}{$ -0.63\substack{+0.21 \\ -0.16}$}
\newcommand{\SMAA}{$0.02603 \pm 0.00040$}
\newcommand{\INCLINATIONA}{$88.3 \substack{+1.2 \\ -1.7}$}
\newcommand{\DURATIONA}{$2.893 \pm 0.022$}
\newcommand{\SecondaryRadiusA}{$1.724 \pm 0.042$}
\newcommand{\SIGMAWA}{$214 \pm 5$}
\newcommand{\SIGMARA}{$16517 \pm 227$}

\newcommand{\PERIODB}{$1.2198769 \pm 0.0000011$}
\newcommand{\EPOCHB}{$8792.63167 \pm 0.00049$}
\newcommand{\ARSTARB}{$3.551 \substack{+0.029 \\ -0.035}$}
\newcommand{\RPRSB}{$0.10712\substack{+0.00076 \\ -0.00074}$}
\newcommand{\IMPACTB}{$0.070 \substack{+0.073 \\ -0.056} $}
\newcommand{\UPLUSB}{$0.802 \substack{+0.086 \\ -0.091} $}
\newcommand{\UMINUSB}{$ -0.53\substack{+0.29 \\ -0.22} $}
\newcommand{\OMSTARB}{$238 \substack{+17 \\ -22} $}
\newcommand{\ISTARB}{$109 \substack{+15 \\ -12} $}
\newcommand{\SMAB}{$0.02603 \pm 0.00040$}
\newcommand{\INCLINATIONB}{$88.8 \substack{+0.8 \\ -1.2} $}
\newcommand{\DURATIONB}{$2.943 \substack{+0.027 \\ -0.022} $}
\newcommand{\SecondaryRadiusB}{$1.692 \pm 0.039$}
\newcommand{\LAMBDAB}{$-148 \substack{+22 \\ -17} $}
\newcommand{\PSIB}{$139 \substack{+12 \\ -15} $}
\newcommand{\SIGMAWB}{$214 \pm 5$}
\newcommand{\SIGMARB}{$16468 \pm 218$}

\newcommand{\PERIODC}{$1.21987628 \substack{+0.00000095 \\ -0.00000102} $}
\newcommand{\EPOCHC}{$8792.63188 \pm 0.00045$}
\newcommand{\ARSTARC}{$3.512 \substack{+0.022 \\ -0.024} $}
\newcommand{\RPRSC}{$0.10696 \pm 0.00083$}
\newcommand{\IMPACTC}{$0.064 \substack{+0.075 \\ -0.058} $}
\newcommand{\UPLUSC}{$0.737 \substack{+0.090 \\ -0.087} $}
\newcommand{\UMINUSC}{$-0.26 \pm 0.28$}
\newcommand{\OMSTARC}{$201.5$ (fixed)}
\newcommand{\ISTARC}{$100.5 \pm 8.4$}
\newcommand{\SMAC}{$0.02603 \pm 0.00040$}
\newcommand{\INCLINATIONC}{$88.9 \substack{+0.8 \\ -1.2} $}
\newcommand{\DURATIONC}{$2.979 \pm 0.016$}
\newcommand{\SecondaryRadiusC}{$1.689 \pm 0.039$}
\newcommand{\LAMBDAC}{$-111.5$ (fixed)}
\newcommand{\PSIC}{$111.3 \substack{+0.2 \\ -0.7} $}
\newcommand{\SIGMAWC}{$214 \pm 6$}
\newcommand{\SIGMARC}{$16471 \pm 247$}

\begin{document}

\title{CHEOPS observations confirm nodal precession in the WASP-33 system}

\titlerunning{CHEOPS observations of the WASP-33 system}
\authorrunning{A. M. S. Smith, et al.}

\author{
A.~M.~S.~Smith\inst{\ref{inst:1}\thanks{\email{alexis.smith@dlr.de}}}\,$^{\href{https://orcid.org/0000-0002-2386-4341}{\protect\includegraphics[height=0.19cm]{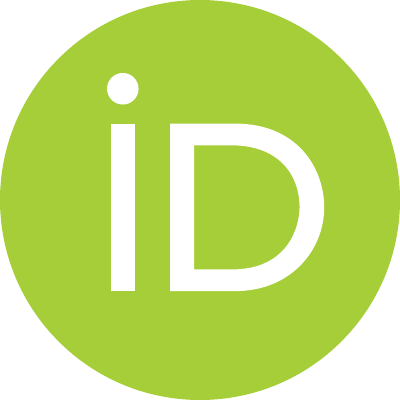}}}$, 
Sz.~Csizmadia\inst{\ref{inst:1}}\,$^{\href{https://orcid.org/0000-0001-6803-9698}{\protect\includegraphics[height=0.19cm]{orcid.pdf}}}$, 
V.~Van~Grootel\inst{\ref{inst:2}}\,$^{\href{https://orcid.org/0000-0003-2144-4316}{\protect\includegraphics[height=0.19cm]{orcid.pdf}}}$, 
M.~Lendl\inst{\ref{inst:3}}\,$^{\href{https://orcid.org/0000-0001-9699-1459}{\protect\includegraphics[height=0.19cm]{orcid.pdf}}}$, 
C.~M.~Persson\inst{\ref{inst:4}}, 
G.~Olofsson\inst{\ref{inst:5}}\,$^{\href{https://orcid.org/0000-0003-3747-7120}{\protect\includegraphics[height=0.19cm]{orcid.pdf}}}$, 
D.~Ehrenreich\inst{\ref{inst:3},\ref{inst:6}}\,$^{\href{https://orcid.org/0000-0001-9704-5405}{\protect\includegraphics[height=0.19cm]{orcid.pdf}}}$, 
M.~N.~G\"unther\inst{\ref{inst:7}}\,$^{\href{https://orcid.org/0000-0002-3164-9086}{\protect\includegraphics[height=0.19cm]{orcid.pdf}}}$, 
A.~Heitzmann\inst{\ref{inst:3}}\,$^{\href{https://orcid.org/0000-0002-8091-7526}{\protect\includegraphics[height=0.19cm]{orcid.pdf}}}$, 
S.~C.~C.~Barros\inst{\ref{inst:8},\ref{inst:9}}\,$^{\href{https://orcid.org/0000-0003-2434-3625}{\protect\includegraphics[height=0.19cm]{orcid.pdf}}}$, 
A.~Bonfanti\inst{\ref{inst:10}}\,$^{\href{https://orcid.org/0000-0002-1916-5935}{\protect\includegraphics[height=0.19cm]{orcid.pdf}}}$, 
A.~Brandeker\inst{\ref{inst:5}}\,$^{\href{https://orcid.org/0000-0002-7201-7536}{\protect\includegraphics[height=0.19cm]{orcid.pdf}}}$, 
J.~Cabrera\inst{\ref{inst:1}}\,$^{\href{https://orcid.org/0000-0001-6653-5487}{\protect\includegraphics[height=0.19cm]{orcid.pdf}}}$, 
O.~D.~S.~Demangeon\inst{\ref{inst:8},\ref{inst:9}}\,$^{\href{https://orcid.org/0000-0001-7918-0355}{\protect\includegraphics[height=0.19cm]{orcid.pdf}}}$, 
L.~Fossati\inst{\ref{inst:10}}\,$^{\href{https://orcid.org/0000-0003-4426-9530}{\protect\includegraphics[height=0.19cm]{orcid.pdf}}}$, 
J.-V.~Harre\inst{\ref{inst:1}}\,$^{\href{https://orcid.org/0000-0001-8935-2472}{\protect\includegraphics[height=0.19cm]{orcid.pdf}}}$,
M.~J.~Hooton\inst{\ref{inst:11}}\,$^{\href{https://orcid.org/0000-0003-0030-332X}{\protect\includegraphics[height=0.19cm]{orcid.pdf}}}$, 
S.~Hoyer\inst{\ref{inst:12}}\,$^{\href{https://orcid.org/0000-0003-3477-2466}{\protect\includegraphics[height=0.19cm]{orcid.pdf}}}$, 
Sz.~Kalman\inst{\ref{inst:13},\ref{inst:14},\ref{inst:15}}, 
S.~Salmon\inst{\ref{inst:3}}\,$^{\href{https://orcid.org/0000-0002-1714-3513}{\protect\includegraphics[height=0.19cm]{orcid.pdf}}}$, 
S.~G.~Sousa\inst{\ref{inst:8}}\,$^{\href{https://orcid.org/0000-0001-9047-2965}{\protect\includegraphics[height=0.19cm]{orcid.pdf}}}$, 
Gy.~M.~Szabó\inst{\ref{inst:16},\ref{inst:14}}\,$^{\href{https://orcid.org/0000-0002-0606-7930}{\protect\includegraphics[height=0.19cm]{orcid.pdf}}}$, 
T.~G.~Wilson\inst{\ref{inst:17}}\,$^{\href{https://orcid.org/0000-0001-8749-1962}{\protect\includegraphics[height=0.19cm]{orcid.pdf}}}$, 
Y.~Alibert\inst{\ref{inst:18},\ref{inst:19}}\,$^{\href{https://orcid.org/0000-0002-4644-8818}{\protect\includegraphics[height=0.19cm]{orcid.pdf}}}$, 
R.~Alonso\inst{\ref{inst:20},\ref{inst:21}}\,$^{\href{https://orcid.org/0000-0001-8462-8126}{\protect\includegraphics[height=0.19cm]{orcid.pdf}}}$, 
J.~Asquier\inst{\ref{inst:7}}, 
T.~B\'arczy\inst{\ref{inst:22}}\,$^{\href{https://orcid.org/0000-0002-7822-4413}{\protect\includegraphics[height=0.19cm]{orcid.pdf}}}$, 
D.~Barrado\inst{\ref{inst:23}}\,$^{\href{https://orcid.org/0000-0002-5971-9242}{\protect\includegraphics[height=0.19cm]{orcid.pdf}}}$, 
W.~Baumjohann\inst{\ref{inst:10}}\,$^{\href{https://orcid.org/0000-0001-6271-0110}{\protect\includegraphics[height=0.19cm]{orcid.pdf}}}$, 
W.~Benz\inst{\ref{inst:19},\ref{inst:18}}\,$^{\href{https://orcid.org/0000-0001-7896-6479}{\protect\includegraphics[height=0.19cm]{orcid.pdf}}}$, 
N.~Billot\inst{\ref{inst:3}}\,$^{\href{https://orcid.org/0000-0003-3429-3836}{\protect\includegraphics[height=0.19cm]{orcid.pdf}}}$, 
L.~Borsato\inst{\ref{inst:24}}\,$^{\href{https://orcid.org/0000-0003-0066-9268}{\protect\includegraphics[height=0.19cm]{orcid.pdf}}}$, 
C.~Broeg\inst{\ref{inst:19},\ref{inst:18}}\,$^{\href{https://orcid.org/0000-0001-5132-2614}{\protect\includegraphics[height=0.19cm]{orcid.pdf}}}$, 
A.~Collier~Cameron\inst{\ref{inst:25}}\,$^{\href{https://orcid.org/0000-0002-8863-7828}{\protect\includegraphics[height=0.19cm]{orcid.pdf}}}$, 
A.~C.~M.~Correia\inst{\ref{inst:26}}\,$^{\href{https://orcid.org/0000-0002-8946-8579}{\protect\includegraphics[height=0.19cm]{orcid.pdf}}}$, 
P.~E.~Cubillos\inst{\ref{inst:10},\ref{inst:27}}, 
M.~B.~Davies\inst{\ref{inst:28}}\,$^{\href{https://orcid.org/0000-0001-6080-1190}{\protect\includegraphics[height=0.19cm]{orcid.pdf}}}$, 
M.~Deleuil\inst{\ref{inst:12}}\,$^{\href{https://orcid.org/0000-0001-6036-0225}{\protect\includegraphics[height=0.19cm]{orcid.pdf}}}$, 
A.~Deline\inst{\ref{inst:3}}, 
B.-O.~Demory\inst{\ref{inst:18},\ref{inst:19}}\,$^{\href{https://orcid.org/0000-0002-9355-5165}{\protect\includegraphics[height=0.19cm]{orcid.pdf}}}$, 
A.~Derekas\inst{\ref{inst:14}}, 
B.~Edwards\inst{\ref{inst:29}}, 
J.~A.~Egger\inst{\ref{inst:19}}\,$^{\href{https://orcid.org/0000-0003-1628-4231}{\protect\includegraphics[height=0.19cm]{orcid.pdf}}}$, 
A.~Erikson\inst{\ref{inst:1}}, 
A.~Fortier\inst{\ref{inst:19},\ref{inst:18}}\,$^{\href{https://orcid.org/0000-0001-8450-3374}{\protect\includegraphics[height=0.19cm]{orcid.pdf}}}$, 
M.~Fridlund\inst{\ref{inst:30},\ref{inst:4}}\,$^{\href{https://orcid.org/0000-0002-0855-8426}{\protect\includegraphics[height=0.19cm]{orcid.pdf}}}$, 
D.~Gandolfi\inst{\ref{inst:31}}\,$^{\href{https://orcid.org/0000-0001-8627-9628}{\protect\includegraphics[height=0.19cm]{orcid.pdf}}}$, 
K.~Gazeas\inst{\ref{inst:32}}\,$^{\href{https://orcid.org/0000-0002-8855-3923}{\protect\includegraphics[height=0.19cm]{orcid.pdf}}}$, 
M.~Gillon\inst{\ref{inst:33}}\,$^{\href{https://orcid.org/0000-0003-1462-7739}{\protect\includegraphics[height=0.19cm]{orcid.pdf}}}$, 
M.~G\"udel\inst{\ref{inst:34}}, 
J.~Hasiba\inst{\ref{inst:10}}, 
Ch.~Helling\inst{\ref{inst:10},\ref{inst:35}}, 
K.~G.~Isaak\inst{\ref{inst:7}}\,$^{\href{https://orcid.org/0000-0001-8585-1717}{\protect\includegraphics[height=0.19cm]{orcid.pdf}}}$, 
L.~L.~Kiss\inst{\ref{inst:36},\ref{inst:37}}, 
J.~Korth\inst{\ref{inst:38}}\,$^{\href{https://orcid.org/0000-0002-0076-6239}{\protect\includegraphics[height=0.19cm]{orcid.pdf}}}$, 
K.~W.~F.~Lam\inst{\ref{inst:1}}\,$^{\href{https://orcid.org/0000-0002-9910-6088}{\protect\includegraphics[height=0.19cm]{orcid.pdf}}}$, 
J.~Laskar\inst{\ref{inst:39}}\,$^{\href{https://orcid.org/0000-0003-2634-789X}{\protect\includegraphics[height=0.19cm]{orcid.pdf}}}$, 
A.~Lecavelier~des~Etangs\inst{\ref{inst:40}}\,$^{\href{https://orcid.org/0000-0002-5637-5253}{\protect\includegraphics[height=0.19cm]{orcid.pdf}}}$, 
D.~Magrin\inst{\ref{inst:24}}\,$^{\href{https://orcid.org/0000-0003-0312-313X}{\protect\includegraphics[height=0.19cm]{orcid.pdf}}}$, 
P.~F.~L.~Maxted\inst{\ref{inst:41}}\,$^{\href{https://orcid.org/0000-0003-3794-1317}{\protect\includegraphics[height=0.19cm]{orcid.pdf}}}$, 
B.~Merín\inst{\ref{inst:42}}\,$^{\href{https://orcid.org/0000-0002-8555-3012}{\protect\includegraphics[height=0.19cm]{orcid.pdf}}}$, 
C.~Mordasini\inst{\ref{inst:19},\ref{inst:18}}, 
V.~Nascimbeni\inst{\ref{inst:24}}\,$^{\href{https://orcid.org/0000-0001-9770-1214}{\protect\includegraphics[height=0.19cm]{orcid.pdf}}}$, 
R.~Ottensamer\inst{\ref{inst:34}}, 
I.~Pagano\inst{\ref{inst:43}}\,$^{\href{https://orcid.org/0000-0001-9573-4928}{\protect\includegraphics[height=0.19cm]{orcid.pdf}}}$, 
E.~Pall\'e\inst{\ref{inst:20},\ref{inst:21}}\,$^{\href{https://orcid.org/0000-0003-0987-1593}{\protect\includegraphics[height=0.19cm]{orcid.pdf}}}$, 
G.~Peter\inst{\ref{inst:44}}\,$^{\href{https://orcid.org/0000-0001-6101-2513}{\protect\includegraphics[height=0.19cm]{orcid.pdf}}}$, 
D.~Piazza\inst{\ref{inst:45}}, 
G.~Piotto\inst{\ref{inst:24},\ref{inst:46}}\,$^{\href{https://orcid.org/0000-0002-9937-6387}{\protect\includegraphics[height=0.19cm]{orcid.pdf}}}$, 
D.~Pollacco\inst{\ref{inst:17}}, 
D.~Queloz\inst{\ref{inst:47},\ref{inst:11}}\,$^{\href{https://orcid.org/0000-0002-3012-0316}{\protect\includegraphics[height=0.19cm]{orcid.pdf}}}$, 
R.~Ragazzoni\inst{\ref{inst:24},\ref{inst:46}}\,$^{\href{https://orcid.org/0000-0002-7697-5555}{\protect\includegraphics[height=0.19cm]{orcid.pdf}}}$, 
N.~Rando\inst{\ref{inst:7}}, 
H.~Rauer\inst{\ref{inst:1},\ref{inst:48}}\,$^{\href{https://orcid.org/0000-0002-6510-1828}{\protect\includegraphics[height=0.19cm]{orcid.pdf}}}$, 
I.~Ribas\inst{\ref{inst:49},\ref{inst:50}}\,$^{\href{https://orcid.org/0000-0002-6689-0312}{\protect\includegraphics[height=0.19cm]{orcid.pdf}}}$, 
N.~C.~Santos\inst{\ref{inst:8},\ref{inst:9}}\,$^{\href{https://orcid.org/0000-0003-4422-2919}{\protect\includegraphics[height=0.19cm]{orcid.pdf}}}$, 
G.~Scandariato\inst{\ref{inst:43}}\,$^{\href{https://orcid.org/0000-0003-2029-0626}{\protect\includegraphics[height=0.19cm]{orcid.pdf}}}$, 
D.~S\'egransan\inst{\ref{inst:3}}\,$^{\href{https://orcid.org/0000-0003-2355-8034}{\protect\includegraphics[height=0.19cm]{orcid.pdf}}}$, 
A.~E.~Simon\inst{\ref{inst:19},\ref{inst:18}}\,$^{\href{https://orcid.org/0000-0001-9773-2600}{\protect\includegraphics[height=0.19cm]{orcid.pdf}}}$, 
M.~Stalport\inst{\ref{inst:2},\ref{inst:33}}, 
S.~Sulis\inst{\ref{inst:12}}\,$^{\href{https://orcid.org/0000-0001-8783-526X}{\protect\includegraphics[height=0.19cm]{orcid.pdf}}}$, 
S.~Udry\inst{\ref{inst:3}}\,$^{\href{https://orcid.org/0000-0001-7576-6236}{\protect\includegraphics[height=0.19cm]{orcid.pdf}}}$, 
S.~Ulmer-Moll\inst{\ref{inst:3},\ref{inst:19},\ref{inst:2}}\,$^{\href{https://orcid.org/0000-0003-2417-7006}{\protect\includegraphics[height=0.19cm]{orcid.pdf}}}$, 
J.~Venturini\inst{\ref{inst:3}}\,$^{\href{https://orcid.org/0000-0001-9527-2903}{\protect\includegraphics[height=0.19cm]{orcid.pdf}}}$, 
E.~Villaver\inst{\ref{inst:20},\ref{inst:21}}, 
V.~Viotto\inst{\ref{inst:24}},
I.~Walter\inst{\ref{inst:51}}\,$^{\href{https://orcid.org/0000-0002-5839-1521}{\protect\includegraphics[height=0.19cm]{orcid.pdf}}}$,
N.~A.~Walton\inst{\ref{inst:52}}\,$^{\href{https://orcid.org/0000-0003-3983-8778}{\protect\includegraphics[height=0.19cm]{orcid.pdf}}}$, and
S.~Wolf\inst{\ref{inst:45}}
}

\institute{
\label{inst:1} Institute of Planetary Research, German Aerospace Center (DLR), Rutherfordstrasse 2, 12489 Berlin, Germany \and
\label{inst:2} Space sciences, Technologies and Astrophysics Research (STAR) Institute, Université de Liège, Allée du 6 Août 19C, 4000 Liège, Belgium \and
\label{inst:3} Observatoire astronomique de l'Université de Genève, Chemin Pegasi 51, 1290 Versoix, Switzerland \and
\label{inst:4} Department of Space, Earth and Environment, Chalmers University of Technology, Onsala Space Observatory, 439 92 Onsala, Sweden \and
\label{inst:5} Department of Astronomy, Stockholm University, AlbaNova University Center, 10691 Stockholm, Sweden \and
\label{inst:6} Centre Vie dans l’Univers, Faculté des sciences, Université de Genève, Quai Ernest-Ansermet 30, 1211 Genève 4, Switzerland \and
\label{inst:7} European Space Agency (ESA), European Space Research and Technology Centre (ESTEC), Keplerlaan 1, 2201 AZ Noordwijk, The Netherlands \and
\label{inst:8} Instituto de Astrofisica e Ciencias do Espaco, Universidade do Porto, CAUP, Rua das Estrelas, 4150-762 Porto, Portugal \and
\label{inst:9} Departamento de Fisica e Astronomia, Faculdade de Ciencias, Universidade do Porto, Rua do Campo Alegre, 4169-007 Porto, Portugal \and
\label{inst:10} Space Research Institute, Austrian Academy of Sciences, Schmiedlstrasse 6, A-8042 Graz, Austria \and
\label{inst:11} Cavendish Laboratory, JJ Thomson Avenue, Cambridge CB3 0HE, UK \and
\label{inst:12} Aix Marseille Univ, CNRS, CNES, LAM, 38 rue Frédéric Joliot-Curie, 13388 Marseille, France \and
\label{inst:13} Konkoly Observatory, Research Centre for Astronomy and Earth Sciences, HUN-REN, MTA Centre of Excellence, Konkoly-Thege Mikl\'os \'ut 15-17., 1121, Hungary \and
\label{inst:14} HUN-REN-ELTE Exoplanet Research Group, Szent Imre h. u. 112., Szombathely, H-9700, Hungary \and
\label{inst:15} ELTE E\"otv\"os Lor\'and University, Doctoral School of Physics, Budapest, P\'azm\'any P\'eter s\'et\'any 1/A, 1117, Hungary \and
\label{inst:16} ELTE Gothard Astrophysical Observatory, 9700 Szombathely, Szent Imre h. u. 112, Hungary \and
\label{inst:17} Department of Physics, University of Warwick, Gibbet Hill Road, Coventry CV4 7AL, United Kingdom \and
\label{inst:18} Center for Space and Habitability, University of Bern, Gesellschaftsstrasse 6, 3012 Bern, Switzerland \and
\label{inst:19} Space Research and Planetary Sciences, Physics Institute, University of Bern, Gesellschaftsstrasse 6, 3012 Bern, Switzerland \and
\label{inst:20} Instituto de Astrofísica de Canarias, Vía Láctea s/n, 38200 La Laguna, Tenerife, Spain \and
\label{inst:21} Departamento de Astrofísica, Universidad de La Laguna, Astrofísico Francisco Sanchez s/n, 38206 La Laguna, Tenerife, Spain \and
\label{inst:22} Admatis, 5. Kandó Kálmán Street, 3534 Miskolc, Hungary \and
\label{inst:23} Depto. de Astrofísica, Centro de Astrobiología (CSIC-INTA), ESAC campus, 28692 Villanueva de la Cañada (Madrid), Spain \and
\label{inst:24} INAF, Osservatorio Astronomico di Padova, Vicolo dell'Osservatorio 5, 35122 Padova, Italy \and
\label{inst:25} Centre for Exoplanet Science, SUPA School of Physics and Astronomy, University of St Andrews, North Haugh, St Andrews KY16 9SS, UK \and
\label{inst:26} CFisUC, Departamento de Física, Universidade de Coimbra, 3004-516 Coimbra, Portugal \and
\label{inst:27} INAF, Osservatorio Astrofisico di Torino, Via Osservatorio, 20, I-10025 Pino Torinese To, Italy \and
\label{inst:28} Centre for Mathematical Sciences, Lund University, Box 118, 221 00 Lund, Sweden \and
\label{inst:29} SRON Netherlands Institute for Space Research, Niels Bohrweg 4, 2333 CA Leiden, Netherlands \and
\label{inst:30} Leiden Observatory, University of Leiden, PO Box 9513, 2300 RA Leiden, The Netherlands \and
\label{inst:31} Dipartimento di Fisica, Università degli Studi di Torino, via Pietro Giuria 1, I-10125, Torino, Italy \and
\label{inst:32} National and Kapodistrian University of Athens, Department of Physics, University Campus, Zografos GR-157 84, Athens, Greece \and
\label{inst:33} Astrobiology Research Unit, Université de Liège, Allée du 6 Août 19C, B-4000 Liège, Belgium \and
\label{inst:34} Department of Astrophysics, University of Vienna, Türkenschanzstrasse 17, 1180 Vienna, Austria \and
\label{inst:35} Institute for Theoretical Physics and Computational Physics, Graz University of Technology, Petersgasse 16, 8010 Graz, Austria \and
\label{inst:36} Konkoly Observatory, Research Centre for Astronomy and Earth Sciences, 1121 Budapest, Konkoly Thege Miklós út 15-17, Hungary \and
\label{inst:37} ELTE E\"otv\"os Lor\'and University, Institute of Physics, P\'azm\'any P\'eter s\'et\'any 1/A, 1117 Budapest, Hungary \and
\label{inst:38} Lund Observatory, Division of Astrophysics, Department of Physics, Lund University, Box 118, 22100 Lund, Sweden \and
\label{inst:39} IMCCE, UMR8028 CNRS, Observatoire de Paris, PSL Univ., Sorbonne Univ., 77 av. Denfert-Rochereau, 75014 Paris, France \and
\label{inst:40} Institut d'astrophysique de Paris, UMR7095 CNRS, Université Pierre \& Marie Curie, 98bis blvd. Arago, 75014 Paris, France \and
\label{inst:41} Astrophysics Group, Lennard Jones Building, Keele University, Staffordshire, ST5 5BG, United Kingdom \and
\label{inst:42} European Space Agency, ESA - European Space Astronomy Centre, Camino Bajo del Castillo s/n, 28692 Villanueva de la Cañada, Madrid, Spain \and
\label{inst:43} INAF, Osservatorio Astrofisico di Catania, Via S. Sofia 78, 95123 Catania, Italy \and
\label{inst:44} Institute of Optical Sensor Systems, German Aerospace Center (DLR), Rutherfordstrasse 2, 12489 Berlin, Germany \and
\label{inst:45} Weltraumforschung und Planetologie, Physikalisches Institut, University of Bern, Gesellschaftsstrasse 6, 3012 Bern, Switzerland \and
\label{inst:46} Dipartimento di Fisica e Astronomia "Galileo Galilei", Università degli Studi di Padova, Vicolo dell'Osservatorio 3, 35122 Padova, Italy \and
\label{inst:47} ETH Zurich, Department of Physics, Wolfgang-Pauli-Strasse 2, CH-8093 Zurich, Switzerland \and
\label{inst:48} Institut fuer Geologische Wissenschaften, Freie Universitaet Berlin, Maltheserstrasse 74-100,12249 Berlin, Germany \and
\label{inst:49} Institut de Ciencies de l'Espai (ICE, CSIC), Campus UAB, Can Magrans s/n, 08193 Bellaterra, Spain \and
\label{inst:50} Institut d'Estudis Espacials de Catalunya (IEEC), 08860 Castelldefels (Barcelona), Spain \and
\label{inst:51} German Aerospace Center (DLR), Institute of Optical Sensor Systems, Rutherfordstraße 2, 12489 Berlin \and
\label{inst:52} Institute of Astronomy, University of Cambridge, Madingley Road, Cambridge, CB3 0HA, United Kingdom
}

   \date{Received 06 September 2024; accepted 04 December 2024}

 
  \abstract
   {}
   {We aim to observe the transits and occultations of \planet, which orbits a rapidly rotating $\delta$ Scuti pulsator, with the goal of measuring the orbital obliquity via the gravity-darkening effect, and constraining the geometric albedo via the occultation depth.}
   {We observed four transits and four occultations with CHEOPS, and employ a variety of techniques to remove the effects of the stellar pulsations from the light curves, as well as the usual CHEOPS systematic effects. We also performed a comprehensive analysis of low-resolution spectral and Gaia data to re-determine the stellar properties of \system.}
   {We measure an orbital obliquity \PSIC~degrees, which is consistent with previous measurements made via Doppler tomography. We also measure the planetary impact parameter, and confirm that this parameter is undergoing rapid secular evolution as a result of nodal precession of the planetary orbit. This precession allows us to determine the second-order fluid Love number of the star, which we find agrees well with the predictions of theoretical stellar models. We are unable to robustly measure a unique value of the occultation depth, and emphasise the need for long-baseline observations to better measure the pulsation periods.}
   {}

\keywords{planets and satellites: dynamical evolution and stability -- planets and satellites: fundamental parameters -- planets and satellites: gaseous planets -- planets and satellites: individual:: WASP-33 b -- stars: individual: WASP-33 -- stars: oscillations (including pulsations)}

\maketitle
%

\section{Introduction}
\label{sec:intro}

\begin{table*}
\caption{Stellar obliquity and transit impact parameter measurements for \planet.}
\begin{tabular}{llllll}
\hline
Epoch & Instrument & $\lambda$ & Impact parameter & Original reference & Re-analysis reference \\
(BJD$_\mathrm{TDB}$ & & (deg)\\
-2450000)\\
\hline
4758.52 & TLS & $-108.8\pm1.0$ & $0.218\pm0.008$ & \cite{wasp33} &\\
4782.92 & McD & $-111.3\substack{+0.76\\-0.77}$ & $0.2398\substack{+0.0062\\-0.0058}$ & \cite{wasp33} & \cite{w33_Watanabe_2022} \\
5173.28 & NOT & $-108.4\pm0.7$ & $0.203\pm0.007$ & \cite{wasp33} &\\
5853.97 & Subaru & $-113.96\pm 0.3$ & $0.1578\pm0.0027$ & \cite{w33_Watanabe_2020} & \cite{w33_Watanabe_2022} \\
6934.77 & McD & $-113.00\pm0.37$ & $0.0845\pm0.0031$ & \cite{w33_Johnson_2015} & \cite{w33_Watanabe_2022} \\
7660.59 & HARPS-N & $-111.39\pm0.23$ & $0.0413\pm0.0019$ & \cite{w33_Borsa_2021} & \cite{w33_Watanabe_2022} \\
7733.79 & McD & $-111.32\substack{+0.49\\-0.47}$ & $0.0432\pm0.0039$ & \cite{w33_Watanabe_2022} & \\
8063.15 & *OAC/MuSCAT & -- & $|b| < 0.132$ & \cite{w33_Watanabe_2022} & \\
8131.46 & HARPS-N & $-111.46\pm0.28$ & $0.0034\substack{+0.0024\\-0.0023}$  & \cite{w33_Borsa_2021} & \cite{w33_Watanabe_2022} \\
8403.49 & *TCS/MuSCAT2 & -- & $|b| < 0.067$ & \cite{w33_Watanabe_2022} & \\
8486.45 & HARPS-N & $-111.64\pm0.28$ & $-0.0272\substack{+0.0020\\-0.0021}$ & \cite{w33_Borsa_2021} & \cite{w33_Watanabe_2022} \\
8804.83 & PEPSI & $-109.29\substack{+0.2\\-0.17}$ & -- & \cite{w33_Cauley_2021} & \\
8845.09 & OAC/HIDES & $-112.24\substack{+0.97\\-1.02}$ & $-0.0592\substack{+0.0066\\-0.0065}$ & \cite{w33_Watanabe_2022} & \\
8792.63 & *TESS & $-94.9\pm21.4$ & $0.177\pm0.039$ & \cite{w33_Kalman_2022} & \\
8792.63 & *TESS & $-109.0\substack{+17.6\\-20.2}$ & $-0.12\pm0.08$ & \cite{w33_Dholakia_2022} &\\
\hline
\\
\end{tabular}
\tablefoot{Original reference refers to the paper where the data were first published. Values are taken from the paper listed under 'Re-analysis reference' where available. All measurements are made by Doppler tomography, except those marked by a *  in the 'Instrument' column, which are photometric. Instrument abbreviations are as follows: TLS: the Coud\'e \'Echelle spectrograph on the 2-m Alfred Jensch telescope at the Th\"uringer Landessternwarte Tautenburg. McD: the Tull spectrograph on the 2.7-m Harlan J. Smith Telescope at McDonald Observatory, Texas. NOT: the FIES spectrograph on the 2.6-m Nordic Optical Telscope, La Palma. HARPS-N: the HARPS-N spectrograph on the 3.6-m Telescopio Nazionale Galileo, La Palma. OAC/MuSCAT: the Multicolor Simultaneous Camera for studying Atmospheres of Transiting exoplanets (MuSCAT) at the Okayama Astro Complex (OAC), Japan. TCS/MuSCAT2: MuSCAT2 at the Telescopio Carlos S\'anchez, Tenerife. PEPSI: Potsdam \'Echelle Polarimetric and Spectroscopic Instrument on the 8.4-m Large Binocular Telescope, Arizona. OAC/HIDES: the HIgh Dispersion \'Echelle Spectrograph at the OAC. TESS: the Transiting Exoplanet Survey Satellite.}
\label{tab:literature2}
\end{table*}

\system is a bright, rapidly rotating A-type star, known to host a hot Jupiter in a 1.22-d orbit \citep{wasp33}. The star exhibits non-radial pulsations of the $\delta$~Sct / $\gamma$~Dor type \citep{herrero, w33_Kalman_2022}. The planet, by virtue of its short orbital period and early-type host star, is one of the hottest-known hot Jupiters \citep{w33_smith}, and has therefore been very well studied, particularly through observations to characterise its atmosphere \cite[e.g.][]{w33_spitzer,w33_von_essen_2015,Cont22_w33}.

The rapid rotation of the host star has a significant impact on the type of observations of the system that can be conducted. With $\vsini = 86$~\kms \citep{wasp33}, extreme line-broadening makes precise radial velocity measurements impossible, but \cite{w33_lehmann} were able to use 248 spectra to derive low-precision radial velocities, sufficient to measure the planetary mass as $2.1\pm0.2$~\mjup. The rapid rotation does, however, enable measurement of the stellar obliquity (spin-orbit angle) via Doppler tomography (DT) \citep{wasp33}.

In the decade or so since the discovery, and first DT measurement of \planet, several further measurements have been made with a variety of spectrographs. These observations enabled the detection \citep{w33_Johnson_2015} and detailed characterisation \citep{w33_Watanabe_2022} of nodal precession in the \system system. This is only the second such planetary detection, after that of Kepler-13A\,b \citep{Szabo12_Kep13}. Two more planets, KELT-9\,b and TOI-1518\,b were recently added to this exclusive list by \cite{w33_stephan22} and \cite{toi1518_watanabe}, respectively. All of these planets are hosted by fast-rotating stars, in orbits significantly misaligned with the stellar rotation. Rapid rotation leads to oblateness, and the more oblate a star, the larger the gravitational quadrupole moment (e.g. \citealt{Dicke70}), and hence the faster the  precession. The misalignment of the planetary orbit doesn't increase the rate of precession, but does act to increase the observed rate of change of parameters such as the transit impact parameter (e.g. \citealt{w33_Watanabe_2022}).

The nodal precession of \planet results in secular variations of the orbital parameters. This is manifested observationally, in particular as a sinusoidal evolution of the impact parameter, $b$, with a period equal to the precession period, determined by \cite{w33_Watanabe_2022} to be around 700~yr. \planet is only expected to exhibit transits (as seen from Earth) for around 20 per cent of this time. Other parameters, such as the sky-projected stellar obliquity, $\lambda$, also vary on the same timescale, but this variation is slower with respect to the measurement precision than for $b$. DT measurements lead to very precise measurements of $b$ that allow the nodal precession to be detected and characterised.

Another effect exhibited by \system is gravity darkening, which is a direct result of the stellar oblateness induced by the rapid rotation. The reduced surface gravity at the stellar equator results in a dark belt in the equatorial region. For planets orbiting in a plane misaligned with the stellar equator, this results in transit light curves with an asymmetric shape \citep{Barnes09}. This phenomenon was also first observed in Kepler-13 \citep{Szabo12_Kep13}, and subsequently in several other systems \citep{Ahlers_koi368,Ahlers_koi89,Ahlers_mascara4}. Observing gravity-darkened transits has the potential not only to measure the obliquity from the light curve shape, but also the stellar inclination with respect to the line-of-sight, \istar. Table~\ref{tab:literature2} lists previously published obliquity and impact parameter measurements for \planet, from both DT measurements and gravity-darkened photometry.

CHEOPS (CHaracterising ExOPlanet Satellite) has observed several systems which exhibit gravity-darkened transits, and the precision of its photometry has enabled constraints on both their obliquities and stellar inclinations \citep{w189_cheops, cheops_mascara1, kelt20_cheops}. These previous observations are summarised in Table~\ref{tab:cheopsGD}. CHEOPS has also observed the occultations of several hot Jupiters, allowing constraints to be placed on their geometric albedos  \citep[e.g.][]{cheops_kelt1,cheops_hd189,cheops_wasp178}. Motivated by observations like these, we observed both transits and occultations of \planet, with the goal of measuring the stellar obliquity via gravity darkening, and the albedo via occultations. In Section~\ref{sec:obs} we describe the new CHEOPS observations of \system, in Section~\ref{sec:stellar_params} we re-determine the stellar parameters of \system, and in Section~\ref{sec:pulsations} we discuss our treatment of the stellar pulsations. In Sections~\ref{sec:transits} and \ref{sec:occultations} we describe our fits to the transits and occultations, respectively. We present our results in Section~\ref{sec:results} and in Section~\ref{sec:discuss} we discuss our results and conclude.

\begin{table*}
\centering
\caption{Previously published observations of gravity-darkened transits with CHEOPS.}
\begin{tabular}{lccccccl}
\hline
Planet & $G$ mag & $n_\mathrm{tr}$ & Resid. RMS (ppm /h) & $\lambda$ (deg.) & $\istar$ (deg) & DT prior on $\lambda$? & Reference \\
\hline
WASP-189\,b & 6.57 & 2 & 10 -- 17 & $86.4\substack{+2.9\\-4.4}$ & $75.5\substack{+3.1\\-2.2}$ & N &\cite{w189_cheops}\\
MASCARA-1\,b & 8.25 & 2 & $\approx 40$ & $-54 \pm 26$ & $29.2\substack{+12.0\\-8.2}$ & N &\cite{cheops_mascara1}\\
KELT-20\,b & 7.59 & 5 & $\approx 22$ & $3.9\pm1.1$ &  $88.9\substack{+18\\-19.8}$ & Y & \cite{kelt20_cheops}\\
\hline
\end{tabular}
\tablefoot{RMS of the residuals to the best fit estimated from the per-point RMS reported for MASCARA-1, and the median absolute deviation reported for KELT-20.
}
\label{tab:cheopsGD}
\end{table*}

\section{CHEOPS observations}
\label{sec:obs}

We observed four transits and four occultations of \planet with ESA's CHEOPS \citep{CHEOPS, Fortier_performance}, as part of the CHEOPS consortium's Guaranteed Time Observerations (GTO) programme. The observing log for these CHEOPS measurements can be found in Table~\ref{tab:obslog}. Each `visit' (transit or occultation observation) lasted for 6 -- 12 hours, covering the full eclipse plus some out-of-eclipse baseline before and afterwards, and comprised a series of exposures, each of 19~s duration. There are gaps in the observations for Earth occultation, stray light, and South Atlantic Anomaly crossings, caused by the low Earth orbit of CHEOPS. With a $G$-band magnitude of 8.1, \system is well-suited to observation by CHEOPS.

\begin{figure*}
\centering
\includegraphics[width=12cm,angle=270]{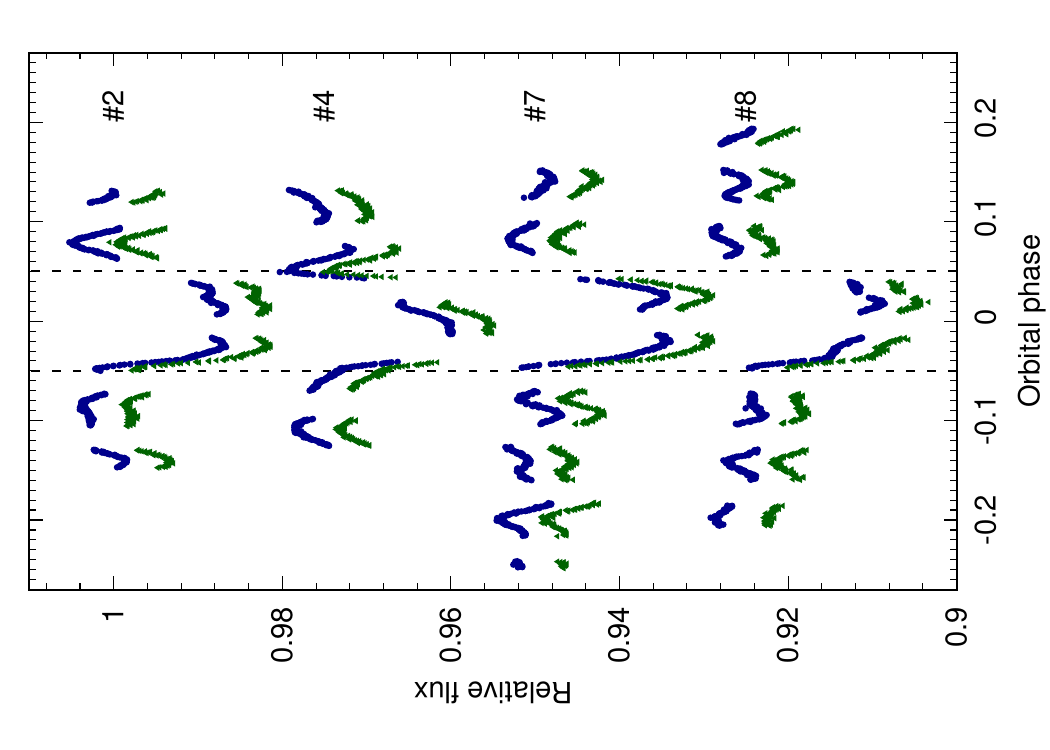}
\includegraphics[width=12cm,angle=270]{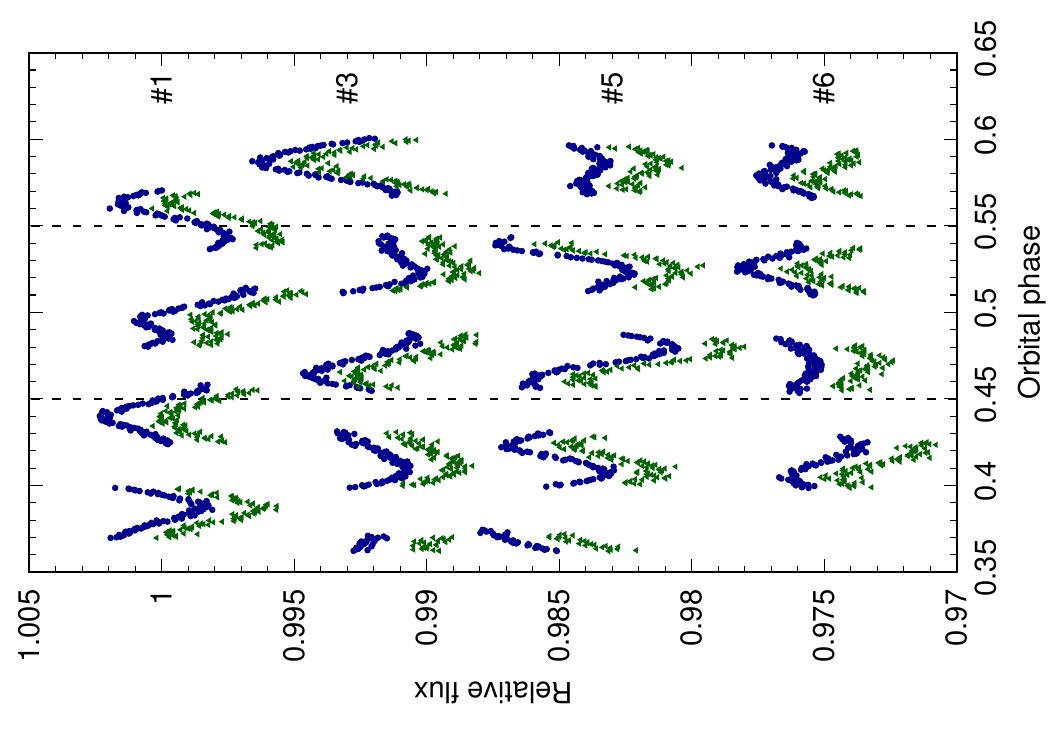}
\caption{CHEOPS photometry of \system covering the transit (left) and occultation (right) of WASP-33\,b. In each case, the {\sc PIPE} light curve (blue circles) is shown with a small vertical offset with respect to the DRP light curve (green triangles). Larger offsets in flux are applied between each transit / occultation for clarity. Dashed vertical lines indicate the beginning and end of each eclipse. The number to the right of each light curve corresponds to the visit number given in Table~\ref{tab:obslog}.}
\label{fig:raw_lcs} 
\end{figure*}

The data were reduced using the standard CHEOPS data reduction pipeline (DRP version 13; \citealt{cheops_drp}), which performs aperture photometry. We also reduced the data with {\sc PIPE}\footnote{\url{https://github.com/alphapsa/PIPE}} (PSF Imagette Photometric Extraction; \citealt{PIPE,Szabo21_AUMic}), an independent data reduction pipeline that relies on PSF fitting. The resulting `raw' light curves from each pipeline are shown in Fig.~\ref{fig:raw_lcs}. In general, the {\sc PIPE} light curves show a lower scatter than those from the DRP, because the `roll angle effect', caused by the rotation of the spacecraft around the optical axis, is greatly reduced in the {\sc PIPE} output, which is less impacted by background stars. We quantified this difference by fitting a fourth order polynomial to a single `chunk' of raw data (that centred on a phase of 0.42 in visit \#5) and measuring the residuals for each of the two light curves. We measured an rms of 39~ppm /hour for the DRP data, and 23~ppm /hour for the PIPE data.

\begin{table*}
\caption{Log of CHEOPS observations of \system.}
\begin{tabular}{lllllll}
\hline
Visit \# & Eclipse & Start date  & Duration & No. of & File key & Efficiency \\
&type& (UTC) & (h) & data points & & (\%) \\\hline
1 & occultation & 2020 Oct 21 21:38 & 5.9 & 377 & CH\_PR100016\_TG007901\_V0200 & 67.2  \\
2 & transit     & 2020 Nov 13 10:45 & 8.1 & 455 & CH\_PR110047\_TG000301\_V0200 & 58.9  \\
3 & occultation & 2020 Nov 18 22:47 & 7.0 & 409 & CH\_PR100016\_TG007902\_V0200 & 61.7  \\
4 & transit     & 2020 Nov 26 21:27 & 7.5 & 439 & CH\_PR110047\_TG000302\_V0200 & 61.4  \\
5 & occultation & 2020 Dec 19 10:44 & 6.9 & 373 & CH\_PR100016\_TG007903\_V0200 & 57.3  \\
6 & occultation & 2020 Dec 24 08:55 & 5.8 & 345 & CH\_PR100016\_TG007904\_V0200 & 62.6  \\
7 & transit     & 2021 Oct 26 23:42 & 11.7 & 662 & CH\_PR110047\_TG001101\_V0200 & 59.6  \\
8 & transit     & 2021 Dec 27 00:50 & 11.7 & 586 & CH\_PR110047\_TG001102\_V0200 & 54.9  \\
\hline
\\
\end{tabular}
\tablefoot{The file keys are unique identifiers for each visit. The visit efficiency is the fraction of each visit for which data was collected.}
\label{tab:obslog}
\end{table*}

\section{Stellar parameters}
\label{sec:stellar_params}

\begin{table}
\caption{Stellar parameters for \system.}
\begin{tabular}{lll}
\hline
Parameter & Value & Source \\
\hline
$G$ & $8.0833 \pm 0.0004$ & Gaia DR3 \\
$G_\mathrm{BP}$ & $8.2089 \pm 0.0014$ & Gaia DR3 \\
$G_\mathrm{RP}$ & $7.8368 \pm 0.0008$ & Gaia DR3 \\
$J$ & $7.581\pm0.021$ & 2MASS \\
$H$ & $7.516\pm0.024$ & 2MASS \\
$K$ & $7.468\pm0.024$ & 2MASS \\
$W1$ & 7.471 & WISE \\
$W2$ & 7.460 & WISE \\
$\varpi$ / mas & $8.2238 \pm 0.0327$ & Gaia DR3 \\
\hline
\teff / K & $7166 \pm 70$ & Spectroscopy (Section~\ref{sec:spec})\\
\logg [cgs] & $4.25 \pm 0.15$ & Spectroscopy (Section~\ref{sec:spec})\\
\rstar/ \rsol & $1.623 \pm 0.036$ & IRFM (Section~\ref{sec:radius})\\
\mstar/ \msol & $1.581\substack{+0.056 \\-0.089}$ & Isochrones (Section~\ref{sec:mass})\\
\agestar / Gyr & $0.6\substack{+0.4 \\-0.3}$ & Isochrones (Section~\ref{sec:mass})\\
\hline\\
\end{tabular}
\tablefoot{
Gaia DR3: \cite{GaiaDR3}. 2MASS: \cite{2MASS}. WISE: \cite{Wright2010}
}
\label{tab:stellar}
\end{table}

\subsection{Spectral parameters}
\label{sec:spec}

\begin{figure}
\centering
\includegraphics[width=\columnwidth]{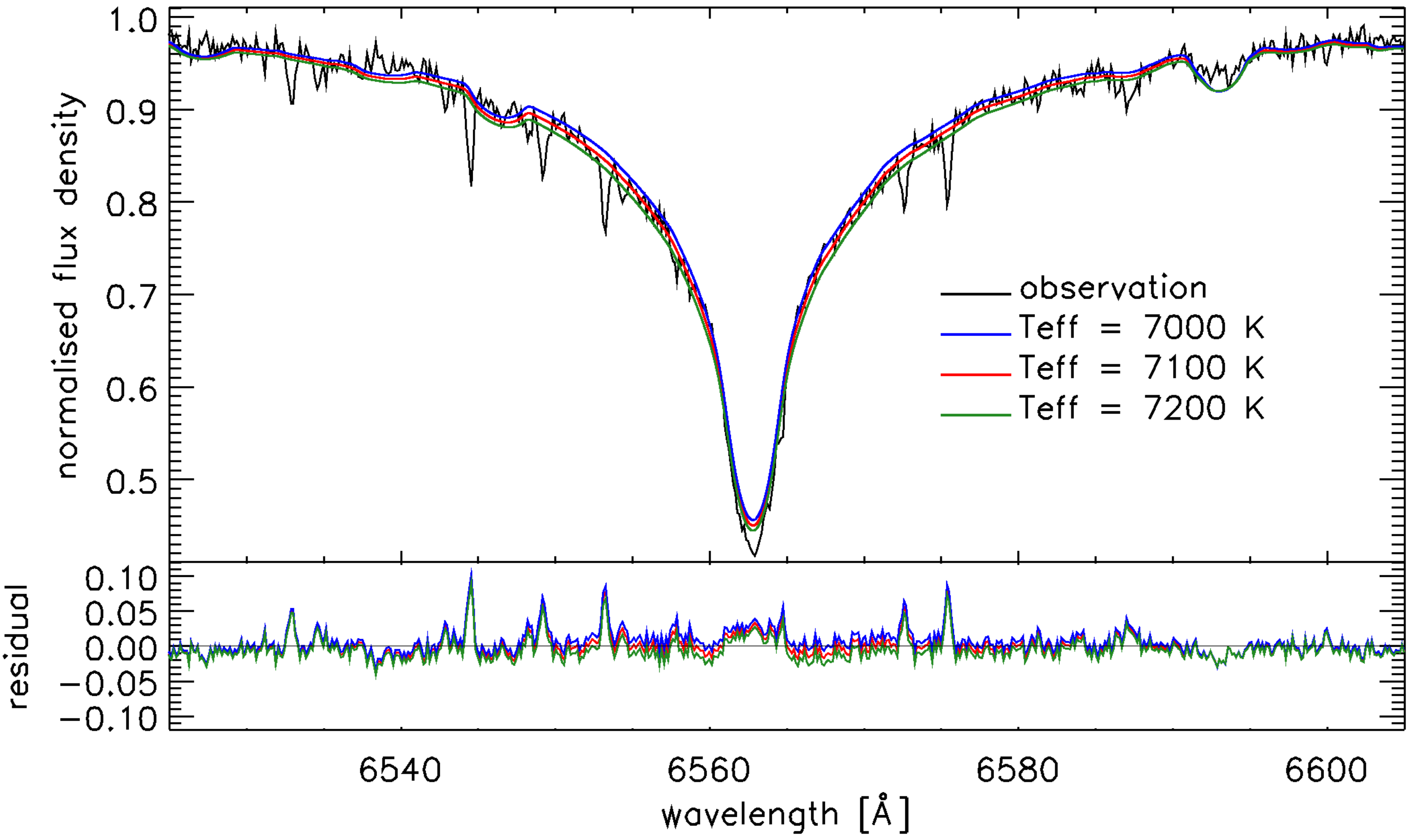}
\caption{Observed H$\alpha$ line profile (black) compared to synthetic spectra computed for \teff values of 7000 (blue), 7100 (red), and 7200\,K (green). The lower panel shows the residuals to the three fits.}
\label{fig:Halpha} 
\end{figure}

We derived the stellar effective temperatute, \teff, using two spectra of WASP-33. First, we used a spectrum with $R$\,$\sim$\,15\,000, which covers the H$\alpha$ line (647--671\,nm), obtained with the 1.8-m telescope of the Dominion Astrophysical Observatory (DAO), Canada on 2011 September 02 \citep[see][for details on the spectral reduction and extraction procedure]{zwintz2013}. The spectrum has a signal-to-noise ratio (S/N) per pixel of 118 at 660~nm. Compared to high-resolution echelle spectra, lower resolution spectra allow a better control of the normalisation of the hydrogen lines, therefore decreasing systematics. To estimate the stellar \teff value from the H$\alpha$ line, we compared the observed spectrum with synthetic spectra calculated with {\sc synth3} \citep{kochukhov2007} on the basis of stellar atmosphere models computed with LLmodels \citep{shulyak2004}. At the temperature of WASP-33, the wings of the Balmer lines are sensitive to \teff variations, while variations in the other parameters (e.g. $\log{g}$, metallicity) play a significantly lesser role \citep[e.g.][]{fuhrmann93}. Fig.~\ref{fig:Halpha} shows a comparison of the DAO spectrum with synthetic spectra computed for \teff values of 7000, 7100, and 7200\,K. The synthetic spectrum for 7100~K fits the observed spectrum better than the other two models. In Fig.~\ref{fig:Halpha}, the observed line core is deeper than the models. This is a result of non-local thermodynamic equilibrium effects that are not included in the calculation of the synthetic spectra. The deep narrow lines in the observed spectrum are of telluric origin.

We also analysed a high-resolution spectrum, consisting of around 350 individual HARPS-N spectra, which has a combined S/N of $\sim 1500$. We model this spectrum with the software Spectroscopy Made Easy\footnote{\url{http://www.stsci.edu/~valenti/sme.html}} 
 \citep[{\tt SME};][]{Valenti1996, pv2017} and the Atlas~12 \citep{Kurucz2013} atmosphere model. Based on  modelling of the H$\alpha$ line wings for \teff and the Ca~I triplet at 610.2~nm, 612.2~nm, and 616.2~nm~for $\log\,g_\star$, we obtain $\teff = 7166 \pm 70$~K (consistent with the result from the low-resolution spectrum, above) and $\log\,g_\star = 4.25 \pm 0.15$ with a fixed   iron abundance and \vsini~of 0.1 and 90~\kms, respectively.

\subsection{Radius}
\label{sec:radius}

To estimate the stellar radius of WASP-33 we used a MCMC modified infrared flux method \citep{Blackwell1977,Schanche2020}. By comparing synthetic photometry, computed from constructed spectral energy distributions (SEDs) using stellar atmospheric models \citep{Castelli2003} and our stellar spectral parameters as priors, to observed broadband photometry in the following bandpasses: \textit{Gaia} $G$, $G_\mathrm{BP}$, and $G_\mathrm{RP}$, 2MASS $J$, $H$, and $K$, and \textit{WISE} $W1$ and $W2$ \citep{2MASS,Wright2010,GaiaDR3}, we derive the stellar bolometric flux. We utilised the Stefan-Boltzmann law to convert the bolometric flux to stellar effective temperature and angular diameter. Our stellar radius is computed using this angular diameter and the offset-corrected \textit{Gaia} parallax \citep{Lindegren2021}.

\subsection{Mass and age}
\label{sec:mass}

We derived the stellar mass \mstar and age \agestar via two different stellar evolutionary models after inputting \teff, [Fe/H], and \rstar along with their errors. In detail, we used the isochrone placement routine \citep{bonfanti2015,bonfanti2016} that interpolates the set of input parameters within pre-computed grids of PARSEC\footnote{\textsl{PA}dova \& T\textsl{R}ieste \textsl{S}tellar \textsl{E}volutionary \textsl{C}ode: \url{http://stev.oapd.inaf.it/cgi-bin/cmd}} v1.2S \citep{marigo2017} isochrones and tracks to compute a first pair of mass and age estimates. A second pair, instead, was derived by the CLES \citep[Code Liègeois d'Évolution Stellaire;][]{scuflaire2008} code, which generates the best-fit stellar evolutionary track based on the input parameters following a Levenberg-Marquadt minimisation scheme \citep{salmon2021}. After checking the consistency between the two respective pairs of outcomes through the $\chi^2$-based criterion as described in \citet{bonfanti2021}, we merged (i.e. we summed) the mass and age distributions and obtained $M_{\star}=1.581_{-0.089}^{+0.056}\,M_{\odot}$ and $\agestar=0.6_{-0.3}^{+0.4}$ Gyr; see \citet{bonfanti2021} for further details.

\section{Stellar pulsations}
\label{sec:pulsations}

As expected, given the hybrid $\gamma$~Dor / $\delta$~Sct \citep{w33_Kalman_2022} nature of \system, the light curves exhibit pulsation-induced variability at a significant amplitude (Fig.~\ref{fig:raw_lcs}). These pulsations must be accounted for when fitting the light curves, particularly the occultation light curves, where the amplitude of the pulsations is significantly larger than the occultation depth. We explored several different approaches to fitting for these pulsations, which are described below.

\subsection{Measuring pulsation frequencies from the CHEOPS data}

We used both {\sc Period04} \citep{period04}, which uses a fast Fourier transform, and Lomb-Scargle periodograms (LSP) as implemented in {\sc astropy} \citep{astropy1, astropy2, astropy3} to measure the frequencies of pulsations from the CHEOPS data themselves. This approach is similar to that performed in earlier studies of the system, such as \cite{w33_smith}. This method is limited by the short baseline of the CHEOPS observations, and also by the gaps in the light curves.

For the occultation data, we also tried determining the frequencies present in each individual CHEOPS visit, using both the aforementioned methods. From the LSP, for a false alarm probability of 0.1 per cent, we calculated 13, 13, 14, and 10 significant frequencies for visits 1, 3, 5, and 6, respectively.

\subsection{Pulsation frequencies measured from TESS data}
\label{sec:puls_tess}

An alternative to measuring the pulsation frequencies directly from the CHEOPS data, is to measure the frequencies from the TESS light curves of \system, which have a much longer baseline. TESS observed \system in sectors 18 (late 2019) and 58 (late 2022). The S18 data were analysed by \cite{w33_vonEssen_2020}, who measured 29 distinct frequencies using {\sc Period04}. A similar approach was employed by \cite{w33_Kalman_2022}, who also analysed the TESS S18 data, producing a list of frequencies slightly different to that of \cite{w33_vonEssen_2020} (Sz.~K\'alm\'an, priv. comm.). 

We performed a third measurement of the pulsation frequencies in the TESS data, by applying the prewhitening technique \citep{1975Ap&SS..36..137D} to the LSP of S18 data with the dedicated software \texttt{FELIX} \citep{2010A&A...516L...6C,Zong2016}. 

The pre-whitening technique consists of subtracting sequentially from the light curve each periodic variation spotted above a given level of S/N. That is, \texttt{FELIX} identifies in the LSP of the light curve the frequency, phase, and amplitude of the highest-amplitude peak, which are used as initial guesses in a subsequent nonlinear least square (NLLS) fit of a cosine wave in time domain using the Levenberg-Marquardt algorithm. The fitted wave of derived frequency, amplitude, and phase is then subtracted from the light curve. The operation was repeated as long as there was a peak above the pre-defined threshold. From the extraction of the second peak and beyond, the model is reevaluated at each step by a simultaneous fit of all the extracted peaks to reevaluate their frequency, phase, and amplitude. The S/N=1 level -- the noise -- is defined locally as the median of the points within a gliding window (centered on each point of the LSP) of 300 times the resolution of the data. This median is re-evaluated at each step of the pre-whitening, that is each time a peak is removed. The minimum significance that we allow for identifying a peak to remove is 4$\sigma$, that is a false alarm probability of $3.2\times10^{-5}$ (99.99994\% probability that the signal is real). This 4$\sigma$ significance level can be converted into an S/N=$x$ level. To determine the value of $x$, we used the approach developed in \citet{Zong2016}: using the same time sampling and the same window as in the TESS light curves, we simulated 10\,000 pure Gaussian white noise light curves. For a given S/N threshold we then searched for the number of times that at least one peak in the LSP of these artificial light curves (that are by construction just noise) happen to be above this threshold. We obtained the false alarm probability by dividing by the number of tests (10,000 here). We found that the threshold corresponding to a 4$\sigma$ significance (false alarm probability of $3.2\times10^{-5}$), is S/N=5.0 for TESS S18. 

The LSP of WASP-33 shows a mix of orbital peaks and stellar pulsations. It is standard, as done in \cite{w33_vonEssen_2020}, to first remove the (primary) transits from the light curve, and then perform the frequency extraction on the LSP of the light curve without transits. We found this approach to be non-optimal, with a lot of residuals linked to the orbital signal remaining in the LSP. We then chose to perform a full frequency extraction using the prewhitening technique on the original light curve (including transits). It is then easy to identify the orbital signal (orbital frequency and its numerous harmonics, here up to 77$* f_{\rm orb}$) and stellar pulsations. The original light curve is then cleaned from the stellar pulsations only, leaving a light curve containing planetary signal only. 108 peaks were extracted from the original light curve of S18 down to S/N=5.0. The frequencies, periods, amplitudes and phases of these 108 peaks, with their associated errors, are presented in Table~\ref{listfreq}. 62 of them can be linked to the orbital signal, being equal (within the errors) to the orbital frequency and multiples of it. The 46 remaining peaks are associated to instrumental noise and stellar signal.  Instrumental noise is often present in TESS data below $\sim$10 $\mu$Hz, and we believe here that all peaks below 13.645 $\mu$Hz are actually of instrumental origin. The 35 remaining peaks, between 22.026 and 394.968 $\mu$Hz, are associated to stellar pulsations of the $\delta$-scuti type. We recovered here almost all of the 29 pulsations detected by \cite{w33_vonEssen_2020}. Four pulsation frequencies from \cite{w33_vonEssen_2020} are very close to a multiple of the orbital frequency, and we listed them here as peaks of orbital origin (see last column of Table~\ref{listfreq}). 

We assume that the pulsations observed in TESS and in CHEOPS have the same frequencies, but different amplitudes, since the passbands of the two instruments differ significantly from each other, with TESS being redder than CHEOPS. Furthermore, we do not expect the phases of the pulsations to remain coherent between the epoch of the TESS observation, and our CHEOPS observations. We therefore keep the frequencies determined from TESS fixed in our analysis of the CHEOPS light curves, but fit for the amplitude and phase (or equivalently the amplitudes of a sine and cosine component) of each frequency. In other words, we fit for the $A_i$s and $B_i$s in this equation:
\begin{equation}
    f_\mathrm{pulsation} = \sum_{i=1}^{n_\nu}A_i{\sin({2\pi \nu_i t_j})} + B_i{\cos({2\pi \nu_i t_j})},
\label{eqn:freqs}
\end{equation}
where a total of $n_\nu$ pulsations are fitted. $\nu_i$ is the frequency of the $i$th pulsation signal and $t_j$ is the timestamp of the $j$th light curve point.

\subsection{Fitting the light curve using measured pulsation frequencies}

As described above, we have five different lists of pulsation frequencies: (i) those measured from the CHEOPS data using {\sc Period04} and (ii) using Lomb-Scargle periodograms, as well as those measured from TESS by (iii) \cite{w33_vonEssen_2020}, (iv) \cite{w33_Kalman_2022}, and (v) ourselves. For each of these frequency lists, we tried fitting for the amplitudes by adding a function like Equation~\ref{eqn:freqs} to the light curve model in \tlcm (Transit and Light Curve Modeller; \citealt{tlcm}).

We also tried a simpler approach to fitting the occultation data, by fitting Equation~\ref{eqn:freqs} along with a simple trapezoid model in which only the depth is variable (i.e. with the total duration, durations of ingress and egress, and the phase of mid-occultation fixed). We iteratively fitted each frequency component alongside the occultation depth and subtracted each pulsation term from the light curve in turn. In both the fitting with \tlcm, and this iterative method, we also tried varying the number of frequencies considered, $n_\nu$.

\subsection{Wavelet-only pulsation correction}
\label{sec:wavelets}

Finally, we tried a method which relies on no direct measurement, or prior knowledge of, the pulsation frequencies. We relied on the wavelet method \citep{wavelets} of correlated noise fitting implemented in \tlcm \citep{tlcm, power_wavelets_1}, with no special treatment for stellar pulsations. This technique was found to be valid for analysing pulsating stars by \cite{bokon23}. In the following two sections, we describe in more detail the transit and occultation models fitted to the data, and how the resulting parameters depend on the treatment of the stellar pulsations.

\section{Fitting the transit data}
\label{sec:transits}

We fitted the CHEOPS transit data using the Transit and Light Curve Modeller (\tlcm; \citealt{tlcm}), which is able to model the asymmetric transits of fast-rotating, gravity-darkened stars. \tlcm was previously used to fit the gravity-darkened transits of WASP-189\,b \citep{w189_cheops}, KELT-20\,b \citep{kelt20_cheops}, and HD\,31221\,b \citep{HD31221}, as well as the TESS light curves of \planet \citep{w33_Kalman_2022}.

We fix the gravity-darkening coefficient, $\beta = 0.23$, based on the oblateness of \system; this is the same value used by \cite{w33_Dholakia_2022}. We note, however, that there is some uncertainty on the appropriate value of $\beta$ to use. Observational studies suggest that values of $\beta$ may vary significantly from those predicted by theory, especially for stars with temperatures similar to \system (e.g. \citealt{grav_dark_beta1}). \cite{w33_Kalman_2022} find $\beta=0.58 \pm 0.20$ from a free fit to the TESS light curve of \system, although they also present a solution with $\beta$ fixed to 0.25. A full exploration of $\beta$ is beyond the scope of this paper.

Initially, we fitted for the following parameters: the orbital period (\porb) the epoch of mid transit ($T_0$), the scaled orbital semi-major axis ($a/ \rstar$), the planet to star radius ratio ($\rplanet/\rstar$), the transit impact parameter ($b$), the stellar inclination angle (\istar), the longitude of the ascending node (\Omstar), the two limb-darkening parameters ($u_+$ and $u_-$)\footnote{Here, $u_+ = u_a + u_b$ and $u_- = u_a - u_b$, where $u_a$ and $u_b$ are the linear and quadratic coefficients respectively, of the quadratic limb-darkening law. This formulation is designed to minimise correlations between the coefficients \citep{tlcm}.}, as well as the photometric white noise ($\sigma_\mathrm{w}$) and red noise ($\sigma_\mathrm{r}$) levels.

Since the data are not able to constrain well the value of $\lambda$, we also performed fits where the value of $\lambda = \Omstar - \Omplanet$ (with $\Omplanet = 90\degr$) is fixed to the value determined from Doppler tomography by \cite{w33_Watanabe_2022}, $\lambda = -111.5\degr$. This allows us to measure better the stellar inclination (from the gravity-darkening effect), as well as the transit impact parameter. We also tried fitting using a transit model that does not include the gravity-darkening effect (i.e. the list of fitted parameters above does not contain \Omstar or \istar, and beyond limb-darkening the stellar disc is assumed to be uniformly bright).

\begin{figure}
\centering
\includegraphics[angle=270,width=\columnwidth]{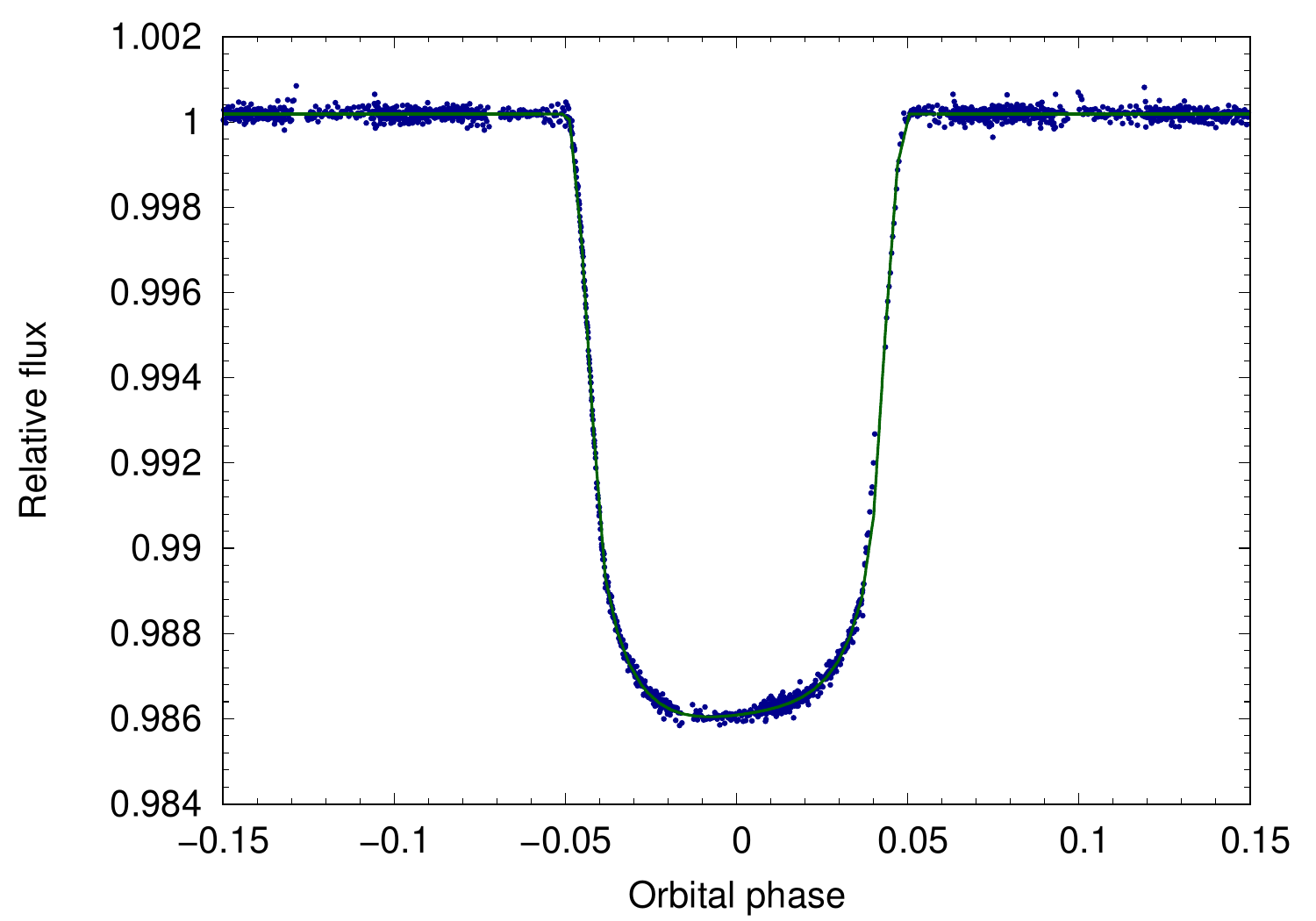}
\includegraphics[angle=270,width=\columnwidth]{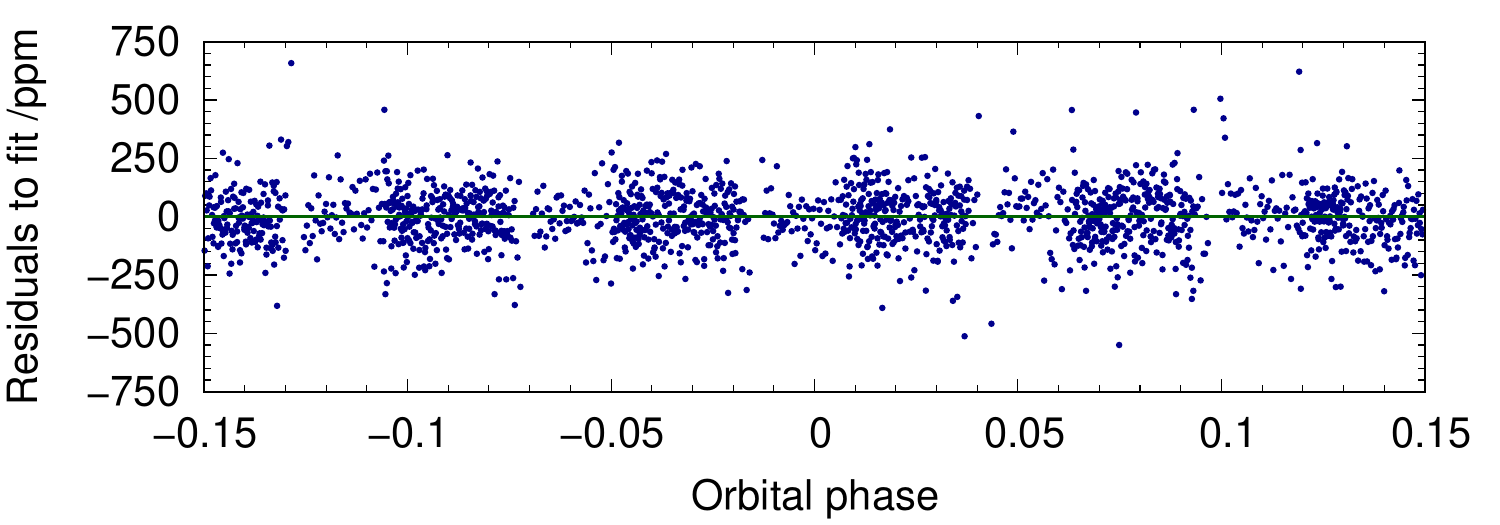}
\includegraphics[angle=270,width=0.97\columnwidth]{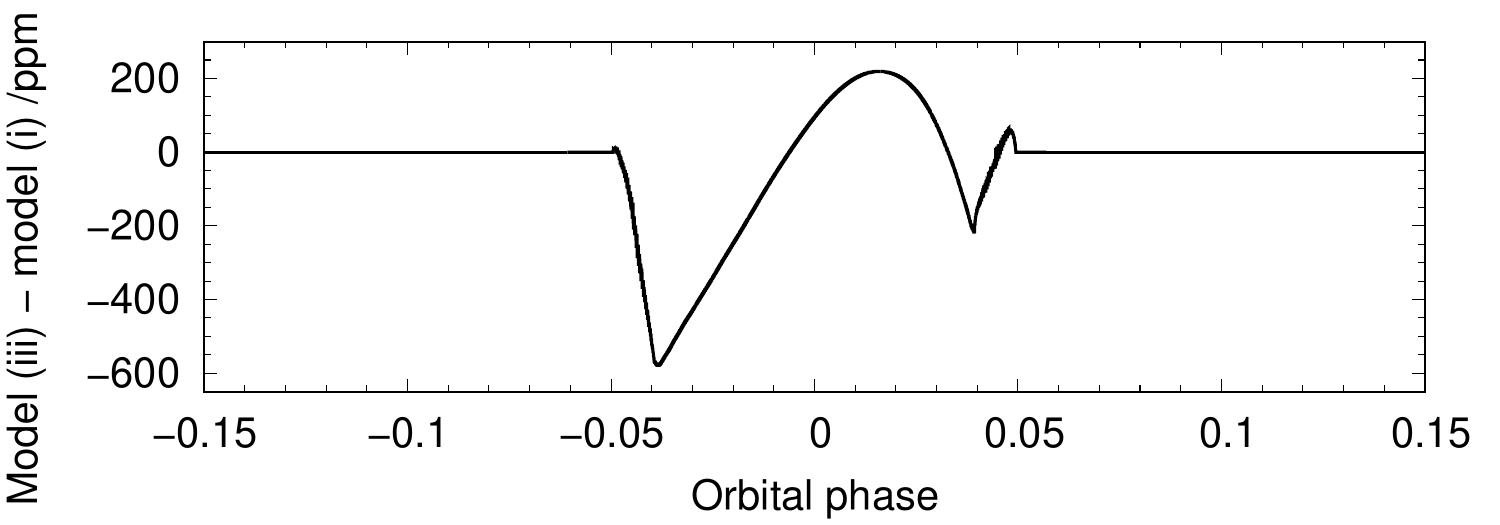}
\caption{Fit to the transit data. Upper panel: Phase-folded transit light curve (blue points), overplotted with the best-fitting model with fixed \Omstar, corresponding to case (iii) of Table~\ref{tab:transit} (green curve). Middle panel: Residuals to the fit shown in the upper panel. Lower panel: The difference between the best-fitting model shown in the upper panel, and the best-fitting model from case (i), where the gravity-darkening phenomenon is not included in the fit.}
\label{fig:transit} 
\end{figure}

We found the results of these fits to be largely insensitive to the details of the pulsation modelling used, as long as sufficient frequencies were considered. The results from the fits where the pulsations were modelled alongside residual systematics using the wavelet method (Section~\ref{sec:wavelets}) are consistent with those where Equation~\ref{eqn:freqs} and a list of frequencies derived from the TESS data were used. When modelling the DRP light curves, we fit for the roll angle effect using a series of sine and cosine terms (see Equation~1 of \citealt{k2_139_cheops}); this is not necessary when using the {\sc PIPE} light curves. We therefore choose to present the results of this wavelet-only fit to the {\sc PIPE} light curves, for the three cases: (i) no gravity darkened model, (ii) gravity darkening with \Omstar a free parameter\footnote{Because of the four-way degeneracy of \istar and \Omstar (see Section~\ref{sec:results_transit} for a detailed discussion of this), we place the following limits on \Omstar and \istar for case (ii): $90\degr < \istar < 180\degr$ and $180\degr < \Omstar < 360\degr$}, and (iii) gravity darkening with \Omstar fixed according to the sky-projected obliquity ($\lambda$) derived from DT (Table~\ref{tab:literature2}). These results are shown in Table~\ref{tab:transit}, and the case (iii) fit to the transit light curves is shown in Fig.~\ref{fig:transit}. The posterior distribution for this fit is shown in Fig.~\ref{fig:corner}, where it can be seen that the largest correlations between parameters are the usual cases: \porb \& $T_0$, $b$ \& $a/ \rstar$, and $u_+$ \& $u_-$.

The resulting parameters for the three cases are all consistent with each other within 2$\sigma$, with the exception of the transit duration, where there is a 3$\sigma$ difference between cases (i) and (iii), corresponding to about 5 minutes. The lower panel of Fig.~\ref{fig:transit} shows the difference between the best-fitting models for cases (i) and (iii), that is between the GD and no-GD cases. The maximum amplitude of this difference is 579~ppm, which can be compared to the amplitudes of the various signals identified in the TESS light curve (Table~\ref{listfreq}); here the orbital harmonics range in amplitude from 34 to 2214~ppm, the instrumental systematics from 154 to 372~ppm, and the stellar pulsations have amplitudes between 49 and 775~ppm. Finally, we used the Bayesian information criterion (BIC) to compare the fits from the three cases (Table~\ref{tab:transit}): the additional complexity of model (iii) is justified.

\begin{table*}
\begin{center}
\caption{Parameters from transit light curve analysis, both with and without gravity darkening (GD).}
\label{tab:transit}
\begin{tabular}{lccc}
\hline
\noalign{\smallskip}
Parameter & (i) no GD & (ii) GD $\lambda$ free & (iii) GD $\lambda$ fixed \\
\noalign{\smallskip}
\hline
\noalign{\smallskip}
Fitted parameters: &&&\\ 
\hline
\noalign{\smallskip}
Orbital period $\porb$ / d  &  \PERIODA  & \PERIODB &\PERIODC\\
Transit epoch $T_0$ / BJD$_\mathrm{TDB}-2450000$  &              \EPOCHA   &\EPOCHB & \EPOCHC\\
Scaled semi-major axis $a/ \rstar$   &      \ARSTARA &\ARSTARB&\ARSTARC\\
Radius ratio $\rplanet/\rstar$   &  \RPRSA & \RPRSB &\RPRSC\\
Transit impact parameter $b$     &     \IMPACTA & \IMPACTB &\IMPACTC\\
Limb-darkening coefficient $u_+$ &   \UPLUSA & \UPLUSB &\UPLUSC\\
Limb-darkening coefficient $u_-$ &   \UMINUSA  & \UMINUSB &\UMINUSC\\
Longitude of ascending node \Omstar & -- & \OMSTARB & \OMSTARC \\
Stellar inclination \istar / deg.  & -- & \ISTARB & \ISTARC \\
White noise level $\sigma_\mathrm{w}$ / ppm & \SIGMAWA & \SIGMAWB & \SIGMAWC \\
Red noise level $\sigma_\mathrm{r}$ / ppm & \SIGMARA & \SIGMARB & \SIGMARC \\
\noalign{\smallskip}
Derived parameters:&&&\\
\hline
\noalign{\smallskip}
Semi-major axis $a$ / au &    \SMAA & \SMAB & \SMAC \\
Orbital inclination angle \iplanet / degrees  & \INCLINATIONA &\INCLINATIONB& \INCLINATIONC\\
Transit Duration $T_{14}$ / h  &  \DURATIONA &\DURATIONB & \DURATIONC\\
Planet radius $R_\mathrm{p} / R_\mathrm{Jup}$  & \SecondaryRadiusA  &\SecondaryRadiusB& \SecondaryRadiusC \\
Sky-projected stellar obliquity $\lambda$ / deg. &  -- & \LAMBDAB & \LAMBDAC \\
True obliquity $\psi$ / deg. & -- & \PSIB & \PSIC \\
Relative Bayesian information criterion (BIC) & 0 & +4.1 & -5.3\\ 
\noalign{\smallskip}
\hline
\end{tabular}
\\
\end{center}
\tablefoot{Tthe best ephemeris for planning future observations is not listed in this table, but rather in Section~\ref{sec:ephemeris}.}
\end{table*}

\section{Fitting the occultation data}
\label{sec:occultations}
\subsection{Fitting}
\label{sec:occultation_fitting}

Since the orbit of \planet is well-determined, and known from previous measurements of the occultation 
to be circular \citep{w33_spitzer}, the only parameter we wish to derive from the CHEOPS occultation data is the occultation depth. However, the amplitude of the stellar pulsations is clearly much larger than the occultation depth -- the occultation is not visible in the raw light curves (Fig.~\ref{fig:raw_lcs}).

We fitted the occultation data using \tlcm, which relies on a physically motivated model, rather than a simple light curve-based approach. We therefore fit only for the ratio of the planetary to stellar intensity (in the CHEOPS bandpass), \Iplanet / \Istar, as well as the parameters needed to characterise the stellar pulsations and residual light curve noise ($\sigma_\mathrm{r}$, $\sigma_\mathrm{w}$). The occultation depth is also reported by \tlcm, and is related to the ratio of intensities by:
\begin{equation}
    \delta_\mathrm{occ} = \frac{\Iplanet}{\Istar} \left(\frac{\rplanet}{\rstar}\right)^2.
\end{equation}

\begin{figure}
\centering
\includegraphics[width=\columnwidth]{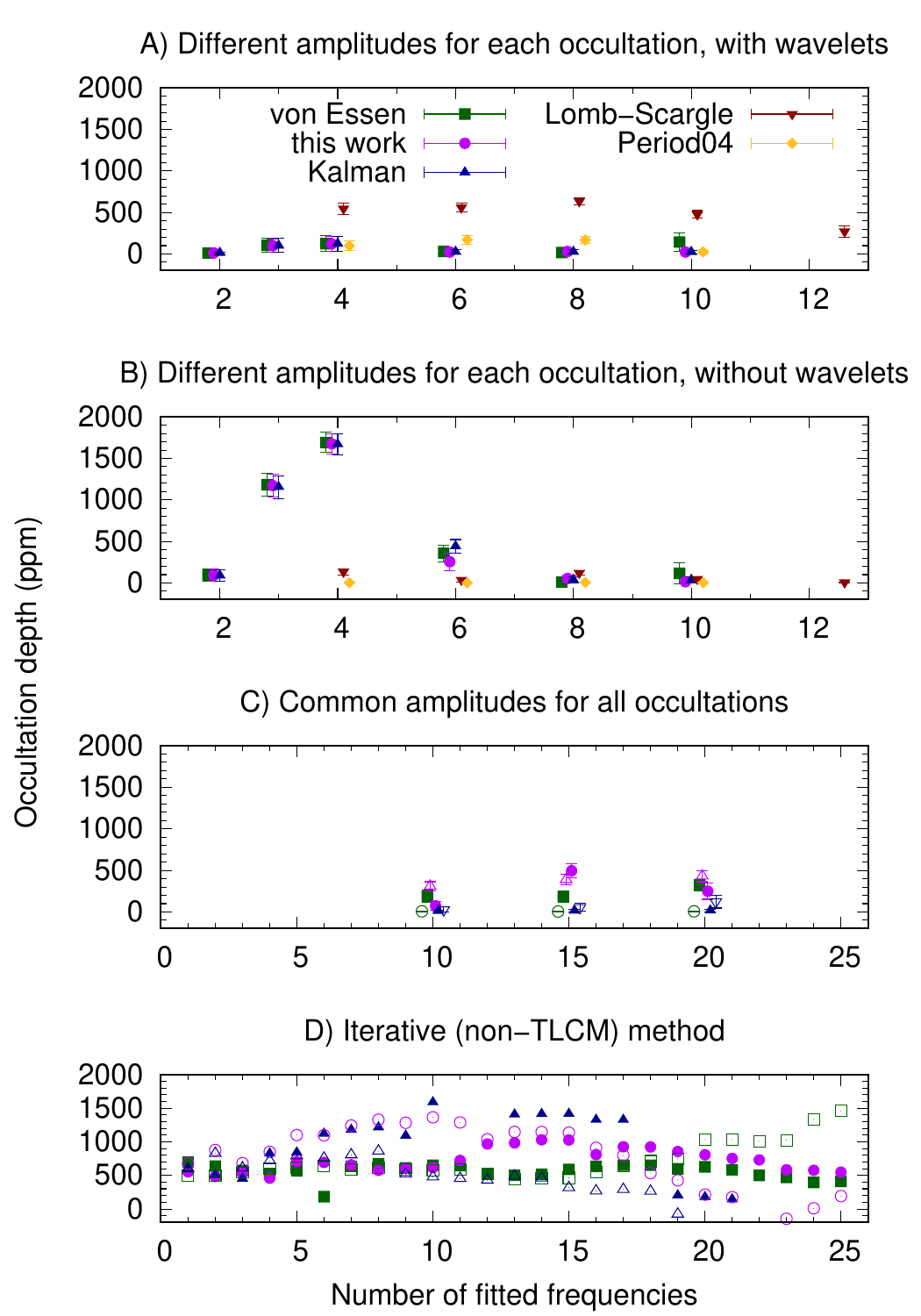}
\caption{Comparison of fitted occultation depths resulting from different approaches to fitting the pulsations. The different colours and symbols correspond to different input lists of pulsation frequencies (see key top right, that applies to all sub panels). A) Different amplitudes were fitted (using TLCM) for each of the four occultations, and residual red noise was fitted with the wavelet method. B) As (A), but without the wavelets. C) A common set of amplitudes was fitted (using TLCM) to all the occultations, with wavelets (solid symbols) and without wavelets (open symbols). D) Here, pulsation signals were fitted and subtracted iteratively, using a simple code, with solid symbols representing a common set of frequencies across all occultations, and open symbols representing unique frequencies for each occultation. In the upper three panels, points are slightly shifted horizontally for readability. The abscissa has different scales in panels A\&B and C\&D.}
\label{fig:occ_depth} 
\end{figure}

All of the methods described in Section~\ref{sec:pulsations} were tried when fitting the occultation data with TLCM. The resulting occultation depth values vary wildly, from zero to more than one thousand parts per million (ppm). Although these most extreme values are probably unphysical, many of our fits resulted in plausible values of the occultation depth (on the order of a few hundred ppm), and seemingly good-looking fits (see Fig.~\ref{fig:occ_apdx} for some examples). However, these values are inconsistent with each other, and do not seem to correlate with either the number of pulsation frequencies fitted, the frequency list used, or the method of fitting (Fig.~\ref{fig:occ_depth}). After performing more than 200 unique fits to the occultation data, we were unfortunately forced to conclude that we are unable to reliably extract an occultation depth from these data. 

We also tried generating synthetic light curves with similar properties to the CHEOPS occultation light curves, with the aim of injecting and recovering an occulation signal. We find that we are unable to reliably recover the injected occultation depth, using some of the same techniques as described above. These tests are described in the following section (\ref{sec:fake}). Our inability to reliably determine the occultation depth means that we are unable to use the CHEOPS data to place any new constraints on the geometric albedo, or other atmospheric parameters of \planet.

\subsection{Injection and recovery tests}
\label{sec:fake}
In order to further investigate the cause of our failure to extract a reliable measurement of the occultation depth from the CHEOPS data, we performed an injection and recovery test. We generated synthetic light curves of the occultation of \planet, using the timestamps from the real CHEOPS occultation data. Fluxes were generated using a simple trapezoidal model for the occultation, with $\delta_\mathrm{occ}$ fixed to 500~ppm. The pulsations characterisation of \cite{w33_vonEssen_2020} were used to generate an artificial pulsation signal, with the values of frequency, amplitude, and phase chosen randomly from normal distributions centred on the best-fitting values, and with standard deviation equal to the uncertainties in their Table~2. White noise was also added to the light curve, with an amplitude of 150~ppm -- the same as that found in the CHEOPS light curves.

\begin{figure}[h]
\centering
\includegraphics[width=\columnwidth]{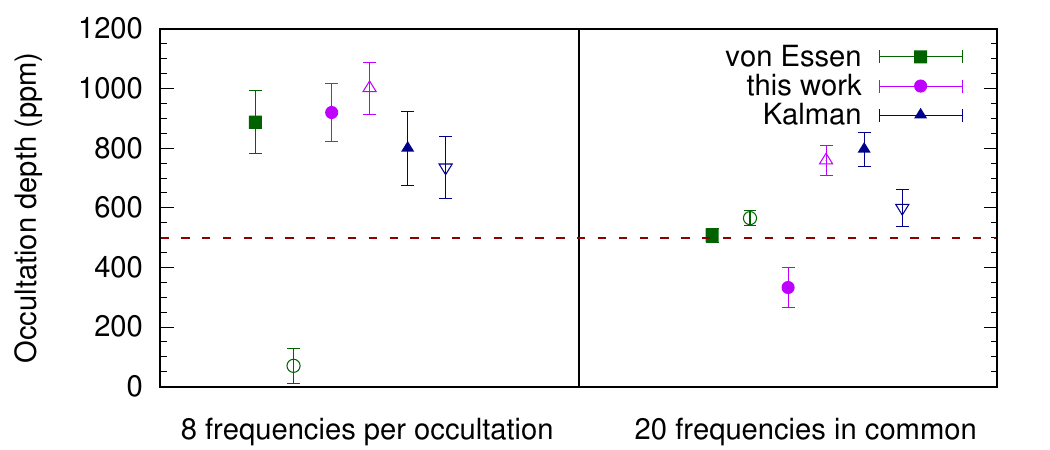}
\caption{Fitted occultation depth for two of the synthetic light curves described in Section~\ref{sec:fake}. We fitted for the pulsations using two approaches: eight frequencies per occultation (left) and twenty frequencies common across all occulations (right). Three different frequency lists were used: from \cite{w33_vonEssen_2020}, \cite{w33_Kalman_2022}, and this work (Section~\ref{sec:puls_tess}). Open and closed symbols refer to the two different randomly generated light curves. The injected occultation depth of 500~ppm is indicated with a dashed red line.}
\label{fig:fake} 
\end{figure}

We then attempted to recover the injected occultation depth, by fitting the synthetic light curves in the same manner as some of our attempts to fit the real data (Section~\ref{sec:occultation_fitting}). The approaches we used here are fitting eight frequencies per occultation, and fitting twenty frequencies common to all occultations. For each approach, we used each of the three TESS pulsation frequencies (Section~\ref{sec:puls_tess}), namely the frequency lists from \cite{w33_vonEssen_2020}, \cite{w33_Kalman_2022}, and our own list, as well as fitting additional noise with the wavelet method. The resulting occultation depths are shown in Fig.~\ref{fig:fake} for two randomly generated synthetic light curves. We fail to reliably recover the injected 500~ppm depth, further justifying our decision not to report an occultation depth for the CHEOPS data.

\section{Results}
\label{sec:results}
\subsection{Transit fit}
\label{sec:results_transit}

Our fits to the transit light curves reveal a clear asymmetry caused by gravity darkening (Fig.~\ref{fig:transit}). Fitting freely for the gravity-darkening parameters results in poorly constrained values of the stellar obliquity and inclination. Fixing the obliquity to the DT value results in a better-constrained value of the stellar inclination \istar = \ISTARC~degrees. This value is consistent with those of \cite{w33_Watanabe_2020} and \cite{w33_Borsa_2021}, who also determine \istar to be slightly larger than $90\degr$, but inconsistent with the values determined by \cite{w33_Watanabe_2022, w33_Dholakia_2022, w33_Kalman_2022} who all find that \istar is less than $90\degr$. These results are largely insensitive to our choice of pulsation treatment.~\\
~\\

There are degeneracies between the angles deduced from gravity darkening. Specifically, transit photometry is unable to distinguish between the following four scenarios \citep{Ahlers_koi368}:

\renewcommand{\labelitemi}{$\circ$}
\begin{itemize}
    \item \istar, $\lambda$
    \item \istar, $180\degr - \lambda$
    \item $180\degr - \istar$, $\lambda$
    \item $180\degr - \istar$, $\lambda + 180\degr$  
\end{itemize}

\begin{figure}
\centering
\includegraphics[width=\columnwidth]{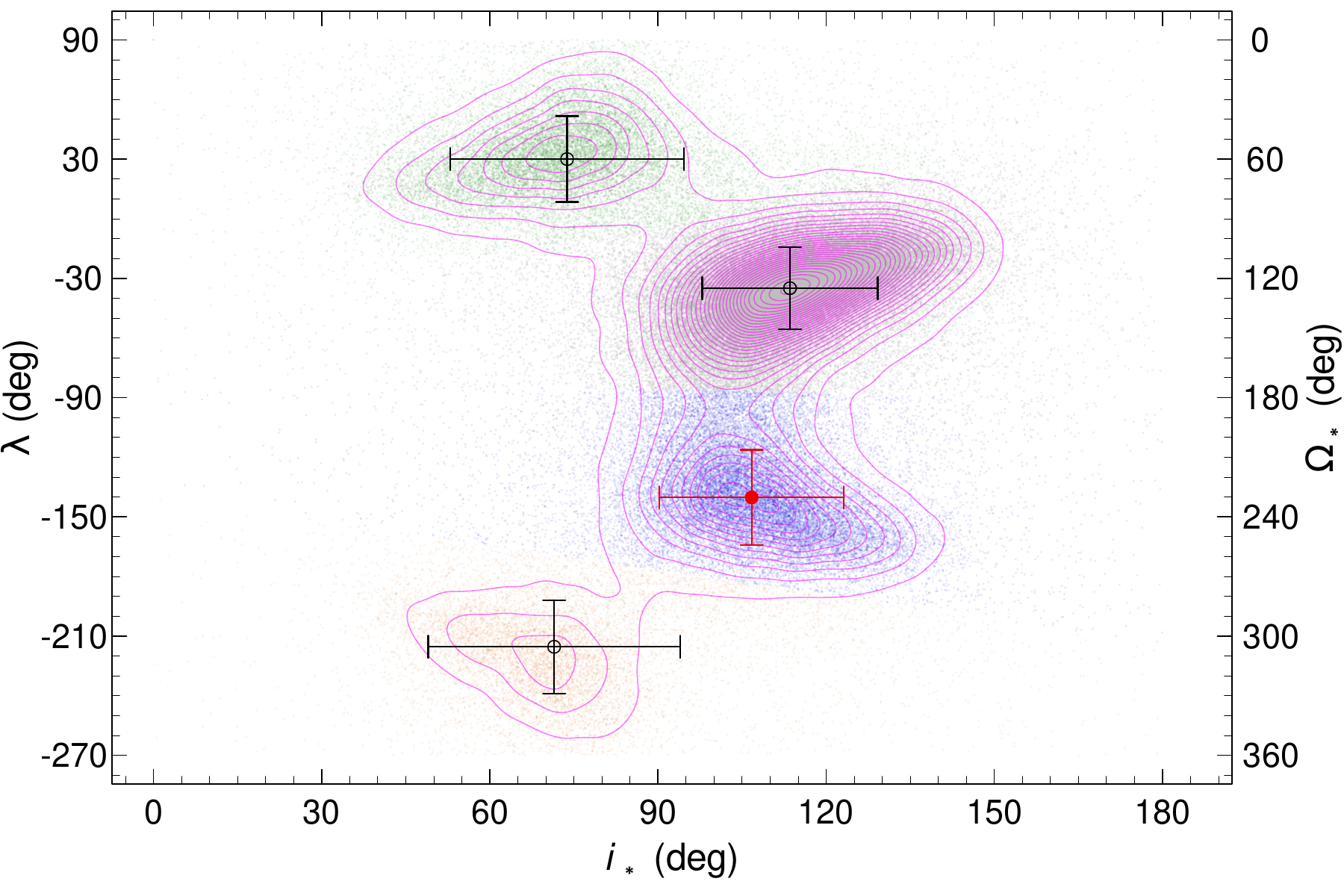}
\caption{Posterior distribution of \Omstar (equivalently $\lambda$) and \istar, for a fit with fitting ranges $0\degr < \istar < 180\degr$ and $0\degr < \Omstar < 360\degr$. A clustering analysis was performed on the posterior, dividing it into four groups, indicated with different colours. The median and $1\sigma$ uncertainties of each group are shown with the open circles and error bars. The red circle is the `correct' solution.  Contours indicating the density of the posterior space are shown with magenta lines.
}
\label{fig:clusters} 
\end{figure}

When we fit for \istar and \Omstar giving a $180\degr$ range for the former, and a $360\degr$ range for the latter, we see all four of the above scenarios represented in the posterior distribution resulting from the \tlcm fit (Fig.~\ref{fig:clusters}). We used a K-means clustering technique \citep{kmeans}, as implemented in the built-in function of the R language, to divide the posterior values into four clusters (Fig.~\ref{fig:clusters}). Although the assignation of posterior values to the four clusters may not be perfect, we note that the results of a fit where \istar and \Omstar were constrained to a single quadrant of the parameter space shown in Fig.~\ref{fig:clusters} ($90\degr < \istar < 180\degr$ and $180\degr < \Omstar < 360\degr$) are virtually identical to the results of the corresponding cluster. The results for case (ii) presented in Table~\ref{tab:transit} are from the fit where \istar and \Omstar are constrained in this way. We also used a two-dimensional kernel density estimation method (implemented in R; \citealt{venables.ripley:modern}) to calculate contours representing the density of the posterior distribution. These contours are shown for 30 levels of density in Fig.~\ref{fig:clusters}.

To test whether four clusters are justified, we computed the silhouette statistic \citep{silhouette} for three, four, and five clusters. We used the python package scikit-learn \citep{scikit-learn} to do this, and found that for three clusters, the value is 0.56, for four clusters 0.56, and for five clusters it is 0.46. The corresponding silhouette plots are shown in Appendix~\ref{sec:silhouette} (Fig.~\ref{fig:silhouette}). These values offer clear evidence against five clusters, but we are unable to choose between three and four clusters using this statistic alone. Nevertheless, we use the geometric argument outlined above, and the fact that the contours indicate that there are four clear local maxima in the density of points in the \Omstar -- \istar plane of the posterior distribution (Fig.~\ref{fig:clusters}), as justifications for our choice of four clusters.

Using the posteriors of our fits, we can compute the true obliquity, $\psi$, using \citep{FabWinn2009},
\begin{equation}
    \cos\psi = \cos \istar \cos \iplanet + \sin\istar \sin\iplanet \cos\lambda.
\end{equation}

For our free fit, we find $\psi$ = \PSIB~deg. and when we fix $\lambda$, $\psi$ = \PSIC~deg., both of which are close to previous determinations, but discrepant at the $2-3\sigma$ level with \citealt{w33_Watanabe_2022} who found $\psi = 108.19\substack{+0.95\\-0.97}$~deg, because our \istar values are inconsistent. Using our values of \istar and \rstar, and $\vsini = 86.0 \pm 0.5$~\kms \citep{wasp33}, we calculate $\prot = 0.94\pm0.03$~d. Alternatively, using the \istar value of \cite{w33_Watanabe_2022}, we obtain $\prot = 0.81\pm0.04$~d.

\subsection{Transit timing and updated ephemeris}
\label{sec:ephemeris}

We also fitted each transit separately to derive the times of each mid-transit. For these fits, we fixed all transit parameters except $T_0$ to the values obtained in our fit with $\lambda$ fixed (case (iii) of Table~\ref{tab:transit}). A similar approach was adopted by \cite{JVH_tidal1}, who derived transit times for several systems observed by CHEOPS in order to search for TTVs caused by tidal orbital decay. 

\begin{figure}
\centering
\includegraphics[width=\columnwidth]{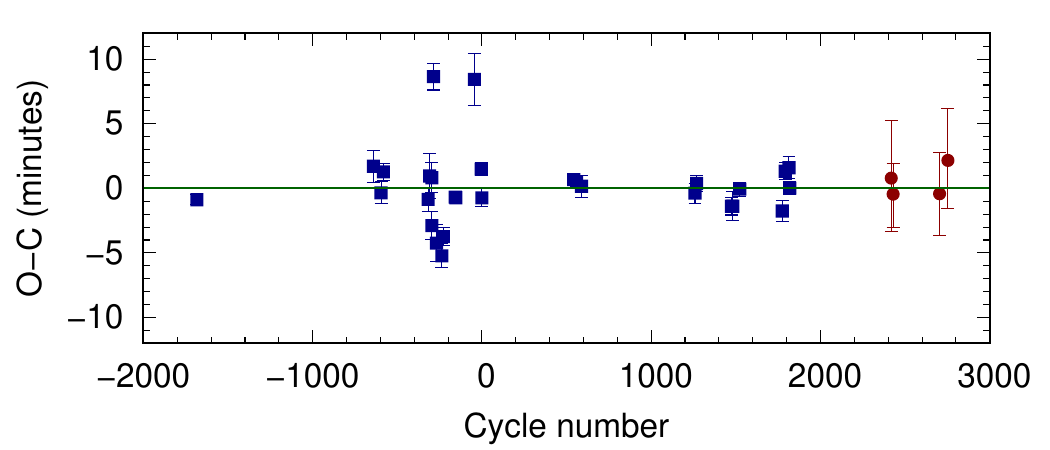}
\caption{Transit timings compared to a linear ephemeris. The archival points collated by \cite{ivshina_winn} are shown with blue squares, relative to their linear ephemeris. The timing of our CHEOPS transits are indicated with red circles.
} 
\label{fig:ttv} 
\end{figure}

The fitted mid-transit times have a mean uncertainty of around 3.5 minutes, and are listed in Table~\ref{tab:ttv}. In Fig.~\ref{fig:ttv}, we plot our transit times alongside those collated and published by \cite{ivshina_winn}, which were originally published by \cite{wasp33, w33_von_essen_2015, w33_Johnson_2015, 2018AJ....155...83Z, 2018AcA....68..371M}. The mid-transit times are compared to the ephemeris published by \cite{ivshina_winn}\footnote{$\porb=1.21987070 \pm 0.00000038$~d, $T_0 = 2456217.48738 \pm 0.00039$ ($\mathrm{BJD_{TDB}}$).}. Our CHEOPS timings have relatively large uncertainties, probably as a result of the data gaps resulting from low-Earth orbit. We also suggest that many of the literature timings have underestimated uncertainties (probably as a result of how the pulsation noise is dealt with). For instance, \cite{power_wavelets_2} found that in the presence of unaccounted for red noise, timing uncertainties can easily be underestimated. The transit timings show no evidence for deviation from a linear ephemeris.

Using the transit timings shown in Fig~\ref{fig:ttv}, we refit for the ephemeris using a simple least squares approach. This yields an ephemeris that is only slightly different to that of \cite{ivshina_winn}, and compatible to within $1 \sigma$. The updated ephemeris is $\porb=1.21987089 \pm 0.00000019$~d, $T_0 = 2456217.48712 \pm 0.00019$ ($\mathrm{BJD_{TDB}}$). This ephemeris predicts transit times with a $1\sigma$ uncertainty less than 90~s for the rest of this decade, and less than 150~s for the entirety of the 2030s.

\begin{table}
\centering
\caption{Fitted times of mid-transit for individual transits of \planet, and the deviations ($O-C$) from the ephemeris presented in \cite{ivshina_winn} [I\&W 2022].}
\begin{tabular}{cccccc}\hline
$E$ &$T_{\rm c} - 2\ 450\ 000$ & $O-C$ \\
(I\&W 2022) &$\mathrm{BJD_{TDB}}$ & d \\
\hline
2418  & $9167.1353\substack{+0.0031\\ -0.0029}$&  $0.0006$ \\
2429  & $9180.5530\substack{+0.0016\\ -0.0018}$&  $-0.0003$ \\
2703  & $9514.7976\substack{+0.0022\\ -0.0022}$&  $-0.0003$ \\
2753  & $9575.7929\substack{+0.0028\\ -0.0026}$&  $0.0015$ \\
\hline
\\
\end{tabular}
\label{tab:ttv}
\end{table}

\subsection{Nodal precession}

\begin{figure}
\centering
\includegraphics[angle=270,width=\columnwidth]{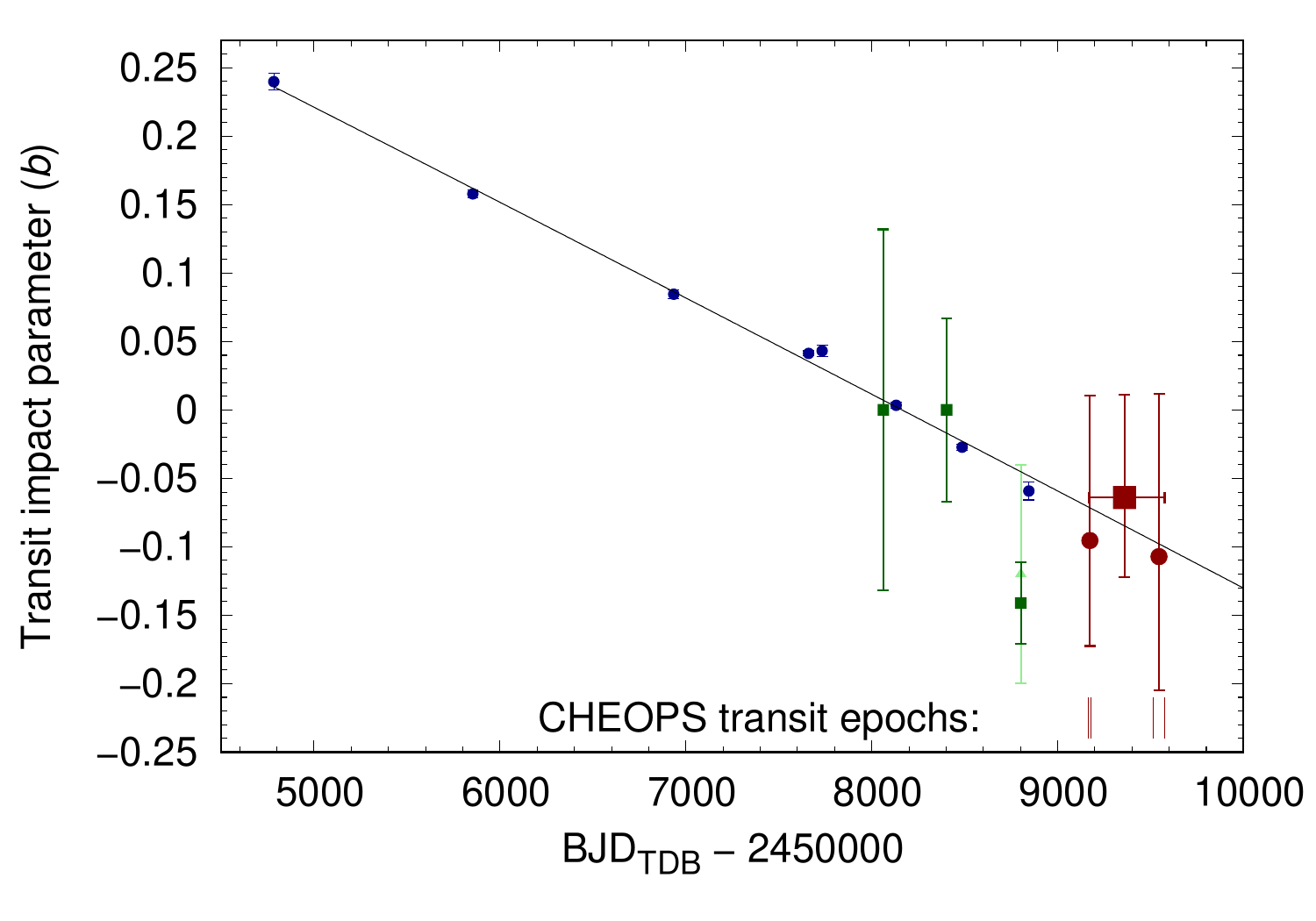}
\caption{Impact parameter as a function of time. Doppler tomographic measurements (from \citealt{w33_Watanabe_2022}) are shown in dark blue, previously published photometric measurements are shown in green, with the two at the same epoch taken from \cite{w33_Dholakia_2022} (light green) and \cite{w33_Kalman_2022} (dark green). Our CHEOPS measurements are shown in dark red, with the result from the fit to all four transits indicated by a square, and the fits to each pair of transits by circles. The epochs of our four CHEOPS transits are shown with short red vertical lines just above the abscissa. The solid black line is the nodal precession model presented in \cite{w33_Watanabe_2022}.} 
\label{fig:b_time} 
\end{figure} 

As discussed in Section~\ref{sec:intro}, nodal precession has previously been detected for \system, with the most recent and comprehensive analysis of this phenomenon published by \cite{w33_Watanabe_2022}. In Fig.~\ref{fig:b_time} we plot the impact parameter derived from our fit to the CHEOPS transits alongside previous such measurements from the literature (Table~\ref{tab:literature2}). We include the data plotted in Fig.~6 of \cite{w33_Watanabe_2022}, namely the eight measurements of $b$ from Doppler tomography, as well as the transit measurements from the MuSCAT and MuSCAT2 instruments. We also include the TESS fits from \cite{w33_Dholakia_2022} and \cite{w33_Kalman_2022}, for which we plot the result for their fit with $\lambda$ fixed. We note that transit light curve fits are not able to distinguish between $+b$ and $-b$\footnote{i.e. between a transit occurring across the northern half of the stellar disc, and one occurring across the southern half.}, but we plot these points with negative $b$, since that is what is expected at these epochs from the DT results. We note that one of these TESS points appears to be a slight outlier ($\sim 2\sigma$), and suggest that this may be the result of the well-known degeneracy between $a/ \rstar$ and $b$ (and perhaps also the gravity-darkening parameters). One of the DT points is similarly discrepant from the model, but with a much smaller error bar, it is less noticeable.

Because the four CHEOPS transits  are spread over more than 400~d, we also performed two additional fits, fitting only the first two transits (which are separated by only 13~d) and only the last two transits (separation: 61~d). The fits were otherwise identical to the $\lambda$-fixed fit described in Section~\ref{sec:transits} The results of these two fits are entirely consistent with that fit, as can be seen in Fig.~\ref{fig:b_time}. The value of $b$ measured from the CHEOPS light curves is also consistent with the nodal precession model of \cite{w33_Watanabe_2022}. The CHEOPS measurements is, however, much less precise than the Doppler tomography measurements. For this reason, we opt not to refit these measurements with a nodal precession model, but simply to show the model fit from \cite{w33_Watanabe_2022} (solid line, Fig.~\ref{fig:b_time}). This model has a precession rate, $\dot{\theta} = 0.507\substack{+0.025\\-0.022}$~deg~yr$^{-1}$, resulting in a stellar gravitational quadrupole moment, $J_2 = \left(1.36\substack{+0.15\\-0.12}\right)\times 10^{-4}$ \citep{w33_Watanabe_2022}. 

\subsection{Stellar Love number}

\begin{figure}
\centering
\includegraphics[angle=0,width=\columnwidth]{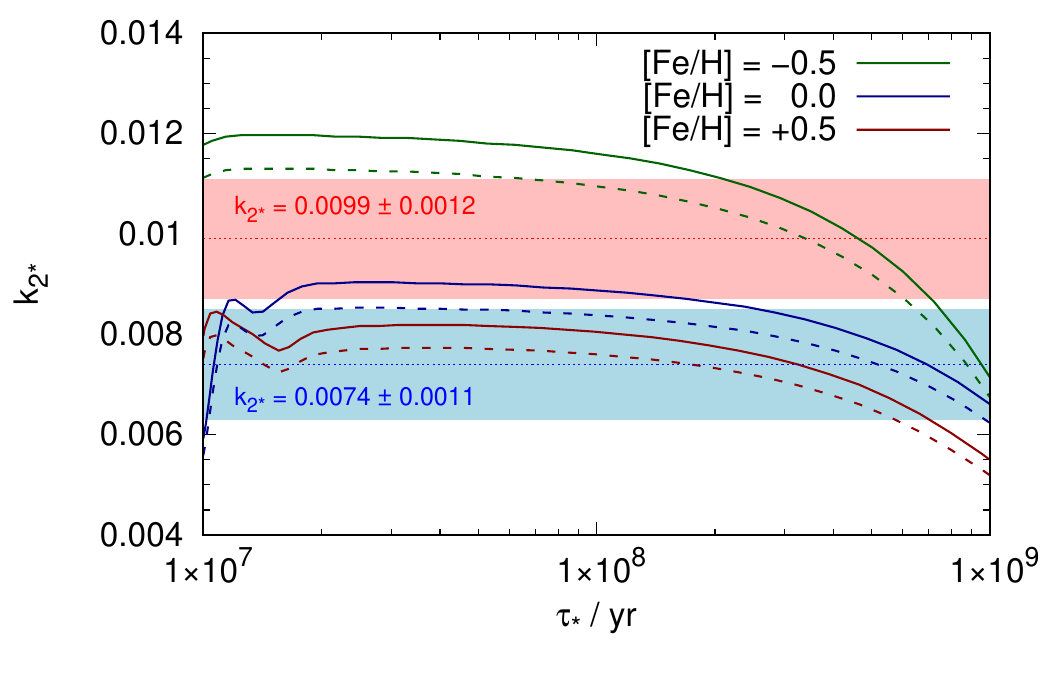}
\caption{Second-order fluid Love number (\ktwostar) of \system as a function of stellar age. Our two calculated values of \ktwostar are indicated with dotted red and blue lines, and their $1\sigma$ confidence intervals with the shaded red and blue areas. Theoretical \ktwostar values from \cite{Claret23} are shown for three different metallicities. In each case the solid curve is for the non-rotating case, and the dashed curve includes a correction factor for the stellar rotation (see \citealt{Claret99}). Both \ktwostar estimates are in good agreement with the theoretical expectations, although given that \feh is probably slightly supersolar, the lower estimate of \ktwostar (corresponding to the \istar value of \citealt{w33_Watanabe_2022}) is in slightly better agreement.
} 
\label{fig:Love} 
\end{figure}

Combining the aforementioned $J_2$ value with the \prot we derived in Section~\ref{sec:results_transit}, we can calculate the second-order fluid Love number of \system, \ktwostar, following Equation~3 of \cite{Ragozzine_Wolf_09}, 
\begin{equation}
\ktwostar = 3J_2 \left(\frac{\Omega_\mathrm{crit}}{\Omega_\mathrm{rot}}\right)^2,
\end{equation}
where $\Omega_\mathrm{crit}^2 = G\mstar/\rstar^3$ is the break-up angular velocity, and $\Omega_\mathrm{rot} = 2\pi/\prot$. We obtain $\ktwostar = 0.0099\pm0.0012$ using our \istar value, and $\ktwostar = 0.0074\pm0.0011$ using the \istar value of \cite{w33_Watanabe_2022}. These can be compared to the theoretical values computed by \cite{Claret23}, noting that the values there must be multiplied by two to give the same quantity as the \ktwostar we calculate above \citep{wasp18_Love}. In Fig.~\ref{fig:Love} we show theoretical values of \ktwostar as a function of stellar age for \mstar = 1.6~\msol. Both of our estimates of \ktwostar are in good agreement with theory, although given that $\feh = 0.1\pm0.1$, the lower estimate of \ktwostar appears to be in better agreement, particularly if \system is at the older end of our age estimate ($\agestar=0.6_{-0.3}^{+0.4}$ Gyr).

\section{Discussion and conclusions}
\label{sec:discuss}

We have confirmed photometrically that the transit impact parameter ($b$) of \planet is undergoing secular evolution, at a rate that is consistent with the nodal precession model presented by \cite{w33_Watanabe_2022}. Although the photometric measurements of $b$ are significantly less precise than those arising from Doppler tomographic fits to transit spectroscopy, our work does demonstrate that such precession can be detected from photometry alone. This has implications for ESA's upcoming PLATO mission \citep{PLATO, PLATO2}, where nodal precession may be detected in several \system-like systems. PLATO's uninterrupted photometric coverage of a large number of bright stars will be unparalleled, allowing secular evolution to be detected in real time.

The long baselines and short cadences of PLATO will also allow the stellar pulsations of stars like \system to be much better characterised. Pulsation characterisation was the limiting factor in our analysis of the CHEOPS light curves. Although we were able to remove the pulsation signals sufficiently well to model the transits, the same cannot be said of our attempts to fit the occultation data. Our sections of out-of-eclipse photometry were just too short (just a few hours) to be able to well characterise the complex spectrum of stellar pulsations at the epoch and in the wavelength band of the eclipses. Using photometry from a different epoch, and passband (i.e. TESS) is far from ideal.

We have also characterised the stellar pulsations of \system from the TESS data, a list of which can be found in Appendix~\ref{sec:pulsation_appendix}. Furthermore, the stellar parameters of \system were redetermined (Section~\ref{sec:stellar_params}), taking into account Gaia data, allowing the most precise and accurate determination of \rstar and \rplanet for this system to date. We determine the stellar radius and mass to be $1.623 \pm 0.036$~\rsol and $1.581\substack{+0.056 \\-0.089}$~\msol, respectively. When combined with the results of our transit fit, we find the planetary radius to be \SecondaryRadiusC~\rjup.

\section*{Data availability}

This article uses data from CHEOPS programmes CH\_PR00016 and CH\_PR00047. The raw and detrended photometric time-series data are available in electronic form at the CDS via anonymous ftp to cdsarc.cds.unistra.fr (130.79.128.5) or via https://cdsarc.cds.unistra.fr/viz-bin/cat/J/A+A/XXX/AYYY.

\begin{acknowledgements}

CHEOPS is an ESA mission in partnership with Switzerland with important contributions to the payload and the ground segment from Austria, Belgium, France, Germany, Hungary, Italy, Portugal, Spain, Sweden, and the United Kingdom. The CHEOPS Consortium would like to gratefully acknowledge the support received by all the agencies, offices, universities, and industries involved. Their flexibility and willingness to explore new approaches were essential to the success of this mission. CHEOPS data analysed in this article will be made available in the CHEOPS mission archive (\url{https://cheops.unige.ch/archive_browser/}). 

Based in part on observations obtained at the Dominion Astrophysical Observatory, Herzberg Astronomy and Astrophysics Research Centre, National Research Council of Canada. We thank David Bohlender for providing the low resolution H$\alpha$ spectrum of the host star.

This research made use of {\sc Astropy} (\url{http://www.astropy.org}), a community-developed core Python package for Astronomy \citep{astropy1, astropy2}, and the {\sc corner} package \citep{corner}.

This work has made use of data from the European Space Agency (ESA) mission
{\it Gaia} (\url{https://www.cosmos.esa.int/gaia}), processed by the {\it Gaia}
Data Processing and Analysis Consortium (DPAC,
\url{https://www.cosmos.esa.int/web/gaia/dpac/consortium}). Funding for the DPAC
has been provided by national institutions, in particular the institutions
participating in the {\it Gaia} Multilateral Agreement.

This research has made use of the NASA Exoplanet Archive, which is operated by the California Institute of Technology, under contract with the National Aeronautics and Space Administration under the Exoplanet Exploration Program.

This publication makes use of The Data \& Analysis Center for Exoplanets (DACE), which is a facility based at the University of Geneva (CH) dedicated to extrasolar planets data visualisation, exchange and analysis. DACE is a platform of the Swiss National Centre of Competence in Research (NCCR) PlanetS, federating the Swiss expertise in Exoplanet research. The DACE platform is available at \url{https://dace.unige.ch}.

The authors gratefully acknowledge the scientific support and HPC resources provided by the German Aerospace Center (DLR). The HPC system CARA is partially funded by "Saxon State Ministry for Economic Affairs, Labour and Transport" and "Federal Ministry for Economic Affairs and Climate Action".

A.M.S.S. and J.-V.H. acknowledge support from the Deutsche Forschungsgemeinschaft grant SM~486/2-1 within the Schwerpunktprogramm SPP 1992 `Exploring the Diversity of Extrasolar Planets'.
V.V.G. is an F.R.S-FNRS Research Associate. 
ML acknowledges support of the Swiss National Science Foundation under grant number PCEFP2\_194576. 
This project has received funding from the Swiss National Science Foundation for project 200021\_200726. It has also been carried out within the framework of the National Centre of Competence in Research PlanetS supported by the Swiss National Science Foundation under grant 51NF40\_205606. The authors acknowledge the financial support of the SNSF. 
MNG is the ESA CHEOPS Project Scientist and Mission Representative, and as such also responsible for the Guest Observers (GO) Programme. MNG does not relay proprietary information between the GO and Guaranteed Time Observation (GTO) Programmes, and does not decide on the definition and target selection of the GTO Programme. 
S.C.C.B. acknowledges support from FCT through FCT contracts nr. IF/01312/2014/CP1215/CT0004. 
ABr was supported by the SNSA. 
This work was supported by FCT - Funda\c{c}\~{a}o para a Ci\^{e}ncia e a Tecnologia through national funds and by FEDER through COMPETE2020 through the research grants UIDB/04434/2020, UIDP/04434/2020, 2022.06962.PTDC. 
O.D.S.D. is supported in the form of work contract (DL 57/2016/CP1364/CT0004) funded by national funds through FCT. 
SH gratefully acknowledges CNES funding through the grant 837319. 
S.G.S. acknowledge support from FCT through FCT contract nr. CEECIND/00826/2018 and POPH/FSE (EC). 
The Portuguese team thanks the Portuguese Space Agency for the provision of financial support in the framework of the PRODEX Programme of the European Space Agency (ESA) under contract number 4000142255. 
GyMSz acknowledges the support of the Hungarian National Research, Development and Innovation Office (NKFIH) grant K-125015, a a PRODEX Experiment Agreement
No. 4000137122, the Lend\"ulet LP2018-7/2021 grant of the Hungarian Academy of Science and the support of the city of Szombathely. 
TWi acknowledges support from the UKSA and the University of Warwick. 
YAl acknowledges support from the Swiss National Science Foundation (SNSF) under grant 200020\_192038.
DBa, EPa, and IRi acknowledge financial support from the Agencia Estatal de Investigaci\'on of the Ministerio de Ciencia e Innovaci\'on MCIN/AEI/10.13039/501100011033 and the ERDF `A way of making Europe' through projects PID2019-107061GB-C61, PID2019-107061GB-C66, PID2021-125627OB-C31, and PID2021-125627OB-C32, from the Centre of Excellence `Severo Ochoa' award to the Instituto de Astrof\'isica de Canarias (CEX2019-000920-S), from the Centre of Excellence `Mar\'ia de Maeztu' award to the Institut de Ci\`encies de l'Espai (CEX2020-001058-M), and from the Generalitat de Catalunya/CERCA programme. 
LBo, VNa, IPa, GPi, RRa, and GSc acknowledge support from CHEOPS ASI-INAF agreement n. 2019-29-HH.0. 
CBr and ASi acknowledge support from the Swiss Space Office through the ESA PRODEX program. 
ACC acknowledges support from STFC consolidated grant number ST/V000861/1, and UKSA grant number ST/X002217/1. 
ACMC acknowledges support from the FCT, Portugal, through the CFisUC projects UIDB/04564/2020 and UIDP/04564/2020, with DOI identifiers 10.54499/UIDB/04564/2020 and 10.54499/UIDP/04564/2020, respectively. 
A.C., A.D., B.E., K.G., and J.K. acknowledge their role as ESA-appointed CHEOPS Science Team Members. 
P.E.C. is funded by the Austrian Science Fund (FWF) Erwin Schroedinger Fellowship, program J4595-N. 
This project was supported by the CNES. 
B.-O. D. acknowledges support from the Swiss State Secretariat for Education, Research and Innovation (SERI) under contract number MB22.00046. 
DG gratefully acknowledges financial support from the CRT foundation under Grant No. 2018.2323 “Gaseous or rocky? Unveiling the nature of small worlds”. 
M.G. is an F.R.S.-FNRS Senior Research Associate. 
CHe acknowledges support from the European Union H2020-MSCA-ITN-2019 under Grant Agreement no. 860470 (CHAMELEON). 
KGI is the ESA CHEOPS Project Scientist and is responsible for the ESA CHEOPS Guest Observers Programme. She does not participate in, or contribute to, the definition of the Guaranteed Time Programme of the CHEOPS mission through which observations described in this paper have been taken, nor to any aspect of target selection for the programme. 
K.W.F.L. was supported by Deutsche Forschungsgemeinschaft grants RA714/14-1 within the DFG Schwerpunkt SPP 1992, Exploring the Diversity of Extrasolar Planets. 
This work was granted access to the HPC resources of MesoPSL financed by the Region Ile de France and the project Equip@Meso (reference ANR-10-EQPX-29-01) of the programme Investissements d'Avenir supervised by the Agence Nationale pour la Recherche. 
PM acknowledges support from STFC research grant number ST/R000638/1. 
This work was also partially supported by a grant from the Simons Foundation (PI Queloz, grant number 327127). 
NCSa acknowledges funding by the European Union (ERC, FIERCE, 101052347). Views and opinions expressed are however those of the author(s) only and do not necessarily reflect those of the European Union or the European Research Council. Neither the European Union nor the granting authority can be held responsible for them. 
JV acknowledges support from the Swiss National Science Foundation (SNSF) under grant PZ00P2\_208945. 
NAW acknowledges UKSA grant ST/R004838/1.\\
~\\
Finally, the authors wish to thank the referee, Dr. Carolina von Essen, whose thorough review and excellent suggestions resulted in significant improvements to this paper.

\end{acknowledgements}

\bibliographystyle{aa}
\bibliography{refs2}


\begin{appendix}
\onecolumn

\section{List of pulsation frequencies detected in the TESS data}
\label{sec:pulsation_appendix}
\footnotesize
\LTcapwidth=\textwidth
\begin{longtable}{lccccccccccl}
\caption{Frequency, period, amplitude, and phase (and associated errors) of the 108 peaks extracted from the LSP of the S18 TESS data, sorted by instrumental, orbital and pulsation origin. The $f_n$ are ranked by decreasing amplitudes. The $F_n$ denotes the \cite{w33_vonEssen_2020} identified stellar pulsations.}\\
\hline\hline
Id. &   & Frequency  & $\sigma_f$ & Period & $\sigma_P$    & Amplitude & $\sigma_A$ & Phase & $\sigma_{\rm Ph}$ & S/N & Comparison to \tabularnewline
    &   & ($\mu$Hz) & ($\mu$Hz)  & (s)    & (s)            & ($\%$)  & ($\%$)    &       &                     &     &    \cite{w33_vonEssen_2020}     \tabularnewline
\hline

 & \tabularnewline
\endfirsthead
\caption{continued.}\\
\hline\hline
Id. &   & Frequency  & $\sigma_f$ & Period & $\sigma_P$    & Amplitude & $\sigma_A$ & Phase & $\sigma_{\rm Ph}$ & S/N & Comments\tabularnewline
    &   & ($\mu$Hz) & ($\mu$Hz)  & (s)    & (s)            & ($\%$)  & ($\%$)    &       &                     &     &         \tabularnewline
\hline

 & \tabularnewline
\endhead

 & \tabularnewline
\hline

\endfoot

 & \tabularnewline
\hline

\endlastfoot
\multicolumn{3}{l}{Instrumental noise} \tabularnewline
$f_{37}$  & & $1.220$   & $0.034$   & $819856$   & $23111$   & $0.0213$   & $0.0027$   & $0.257$   & $0.041$   & $7.8$   &  \tabularnewline
$f_{18}$  & & $2.843$   & $0.020$   & $351786$   & $2424$   & $0.0372$   & $0.0027$   & $0.859$   & $0.023$   & $13.7$   &  \tabularnewline
$f_{43}$  & & $3.835$   & $0.041$   & $260745$   & $2783$   & $0.0178$   & $0.0027$   & $0.722$   & $0.049$   & $6.6$   &  \tabularnewline
$f_{33}$  & & $4.704$   & $0.033$   & $212602$   & $1480$   & $0.0223$   & $0.0027$   & $0.692$   & $0.039$   & $8.2$   & \tabularnewline
$f_{48}$  & & $6.057$   & $0.045$   & $165093$   & $1227$   & $0.0162$   & $0.0027$   & $0.531$   & $0.053$   & $6.0$   & \tabularnewline
$f_{42}$  & & $10.576$   & $0.040$   & $94554$   & $354$   & $0.0179$   & $0.0026$   & $0.176$   & $0.047$   & $6.8$   &  \tabularnewline
$f_{22}$  & & $10.979$   & $0.023$   & $91081$   & $193$   & $0.0303$   & $0.0026$   & $0.132$   & $0.028$   & $11.5$   &  \tabularnewline
$f_{51}$  & & $12.315$   & $0.046$   & $81200$   & $302$   & $0.0154$   & $0.0026$   & $0.136$   & $0.054$   & $5.9$   &  \tabularnewline
$f_{44}$  & & $13.645$   & $0.040$   & $73289$   & $215$   & $0.0175$   & $0.0026$   & $0.092$   & $0.047$   & $6.7$   &  \tabularnewline
\tabularnewline
\multicolumn{3}{l}{Orbital peaks} \tabularnewline
$f_{2}$  & & $9.4836$   & $0.0033$   & $105445$   & $36$   & $0.2200$   & $0.0027$   & $0.8851$   & $0.0039$   & $82.4$   & $f_{orb}$\tabularnewline
$f_{1}$  & & $18.9745$   & $0.0031$   & $52702.3$   & $8.7$   & $0.2214$   & $0.0026$   & $0.2559$   & $0.0037$   & $85.7$   & $2*f_{orb}$ \tabularnewline
$f_{4}$  & & $28.4211$   & $0.0037$   & $35185.2$   & $4.6$   & $0.1794$   & $0.0025$   & $0.6855$   & $0.0044$   & $72.8$   & $3*f_{orb}$ \tabularnewline
$f_{3}$  & & $37.9461$   & $0.0034$   & $26353.1$   & $2.4$   & $0.1831$   & $0.0023$   & $0.0247$   & $0.0041$   & $78.3$   & $4*f_{orb}$ \tabularnewline
$f_{5}$  & & $47.4358$   & $0.0039$   & $21081.1$   & $1.7$   & $0.1538$   & $0.0022$   & $0.3978$   & $0.0046$   & $69.2$   & $5*f_{orb}$ \tabularnewline
$f_{6}$  & & $56.9267$   & $0.0039$   & $17566.5$   & $1.2$   & $0.1454$   & $0.0021$   & $0.7782$   & $0.0046$   & $68.7$   & $6*f_{orb}$ \tabularnewline
$f_{7}$  & & $66.4181$   & $0.0050$   & $15056.1$   & $1.1$   & $0.1089$   & $0.0020$   & $0.1486$   & $0.0059$   & $54.2$   & $7*f_{orb}$ \tabularnewline
$f_{8}$  & & $75.8935$   & $0.0056$   & $13176.37$   & $0.98$   & $0.0925$   & $0.0019$   & $0.5492$   & $0.0067$   & $47.7$   & $8*f_{orb}$ \tabularnewline
$f_{12}$  & & $85.3925$   & $0.0087$   & $11710.6$   & $1.2$   & $0.0569$   & $0.0019$   & $0.907$   & $0.010$   & $30.7$   & $9*f_{orb}$ \tabularnewline
$f_{17}$  & & $94.911$   & $0.013$   & $10536.1$   & $1.4$   & $0.0375$   & $0.0018$   & $0.262$   & $0.015$   & $20.9$   & $10*f_{orb}$ \tabularnewline
$f_{45}$  & & $104.396$   & $0.028$   & $9579.0$   & $2.6$   & $0.0174$   & $0.0018$   & $0.637$   & $0.033$   & $9.7$   & $11*f_{orb}$ \tabularnewline
$f_{11}$  & & $113.9649$   & $0.0081$   & $8774.63$   & $0.63$   & $0.0572$   & $0.0017$   & $0.1921$   & $0.0096$   & $33.0$   & $12*f_{orb}$ \tabularnewline
$f_{31}$  & & $123.558$   & $0.019$   & $8093.4$   & $1.2$   & $0.0238$   & $0.0017$   & $0.567$   & $0.022$   & $14.2$   & $13*f_{orb}$ \tabularnewline
$f_{25}$  & & $132.845$   & $0.016$   & $7527.59$   & $0.88$   & $0.0282$   & $0.0016$   & $0.316$   & $0.018$   & $17.3$   & $14*f_{orb}$ \tabularnewline
$f_{19}$  & & $142.322$   & $0.012$   & $7026.31$   & $0.61$   & $0.0345$   & $0.0016$   & $0.679$   & $0.015$   & $21.7$   & $15*f_{orb}$ \tabularnewline
$f_{15}$  & & $151.8103$   & $0.0099$   & $6587.17$   & $0.43$   & $0.0414$   & $0.0015$   & $0.058$   & $0.012$   & $27.2$   & $16*f_{orb}$ \tabularnewline
$f_{16}$  & & $161.2819$   & $0.0094$   & $6200.32$   & $0.36$   & $0.0411$   & $0.0014$   & $0.454$   & $0.011$   & $28.4$   & $17*f_{orb}$ \tabularnewline
$f_{14}$  & & $170.7736$   & $0.0088$   & $5855.71$   & $0.30$   & $0.0417$   & $0.0014$   & $0.822$   & $0.010$   & $30.4$   & $18*f_{orb}$ \tabularnewline
$f_{21}$  & & $180.264$   & $0.011$   & $5547.41$   & $0.35$   & $0.0316$   & $0.0013$   & $0.203$   & $0.013$   & $23.7$   & $19*f_{orb}$ \tabularnewline
$f_{24}$  & & $189.762$   & $0.012$   & $5269.75$   & $0.33$   & $0.0288$   & $0.0013$   & $0.578$   & $0.014$   & $22.5$   & $20*f_{orb}$ \tabularnewline
$f_{41}$  & & $199.252$   & $0.017$   & $5018.78$   & $0.44$   & $0.0188$   & $0.0012$   & $0.951$   & $0.021$   & $15.5$   & $21*f_{orb}$ \tabularnewline
$f_{73}$  & & $208.717$   & $0.030$   & $4791.18$   & $0.68$   & $0.0105$   & $0.0012$   & $0.385$   & $0.035$   & $9.0$   & $22*f_{orb}$ \tabularnewline
$f_{89}$  & & $227.750$   & $0.045$   & $4390.79$   & $0.86$   & $0.0065$   & $0.0011$   & $0.533$   & $0.053$   & $6.0$   & $24*f_{orb}$ \tabularnewline
$f_{71}$  & & $237.171$   & $0.025$   & $4216.37$   & $0.45$   & $0.0109$   & $0.0010$   & $0.991$   & $0.030$   & $10.7$   & $25*f_{orb}$ \tabularnewline
$f_{40}$  & & $246.641$   & $0.014$   & $4054.48$   & $0.23$   & $0.01880$   & $0.00099$   & $0.396$   & $0.017$   & $19.0$   & $26*f_{orb} (F_{24})$ \tabularnewline
$f_{26}$  & & $256.2286$   & $0.0097$   & $3902.76$   & $0.15$   & $0.02598$   & $0.00094$   & $0.702$   & $0.012$   & $27.6$   &  $27*f_{orb} (F_{20}$)  \tabularnewline
$f_{34}$  & & $265.662$   & $0.011$   & $3764.19$   & $0.16$   & $0.02210$   & $0.00091$   & $0.101$   & $0.013$   & $24.2$   & $28*f_{orb}$ \tabularnewline
$f_{36}$  & & $275.131$   & $0.011$   & $3634.64$   & $0.15$   & $0.02144$   & $0.00088$   & $0.498$   & $0.013$   & $24.4$   & $29*f_{orb}$ \tabularnewline
$f_{39}$  & & $284.643$   & $0.012$   & $3513.17$   & $0.15$   & $0.01925$   & $0.00085$   & $0.873$   & $0.014$   & $22.6$   & $30*f_{orb}$ \tabularnewline
$f_{47}$  & & $294.125$   & $0.013$   & $3399.92$   & $0.15$   & $0.01703$   & $0.00083$   & $0.250$   & $0.016$   & $20.5$   & $31*f_{orb}$ \tabularnewline
$f_{56}$  & & $303.598$   & $0.016$   & $3293.83$   & $0.17$   & $0.01379$   & $0.00081$   & $0.638$   & $0.019$   & $16.9$   & $32*f_{orb}$ \tabularnewline
$f_{86}$  & & $313.078$   & $0.030$   & $3194.09$   & $0.30$   & $0.00721$   & $0.00079$   & $0.014$   & $0.035$   & $9.1$   & $33*f_{orb}$ \tabularnewline
$f_{85}$  & & $332.033$   & $0.028$   & $3011.74$   & $0.25$   & $0.00738$   & $0.00076$   & $0.501$   & $0.033$   & $9.7$   & $35*f_{orb} (F_{22}$) \tabularnewline
$f_{88}$  & & $341.586$   & $0.029$   & $2927.52$   & $0.25$   & $0.00690$   & $0.00075$   & $0.604$   & $0.035$   & $9.2$   & $36*f_{orb}$ \tabularnewline
$f_{83}$  & & $351.048$   & $0.025$   & $2848.61$   & $0.21$   & $0.00787$   & $0.00074$   & $0.032$   & $0.030$   & $10.6$   & $37*f_{orb}$ ($F_{29}$) \tabularnewline
$f_{65}$  & & $360.521$   & $0.017$   & $2773.77$   & $0.13$   & $0.01187$   & $0.00074$   & $0.415$   & $0.020$   & $16.1$   & $38*f_{orb}$ \tabularnewline
$f_{59}$  & & $370.038$   & $0.015$   & $2702.42$   & $0.11$   & $0.01333$   & $0.00073$   & $0.761$   & $0.017$   & $18.4$   & $39*f_{orb}$ \tabularnewline
$f_{53}$  & & $379.519$   & $0.013$   & $2634.917$   & $0.091$   & $0.01480$   & $0.00072$   & $0.154$   & $0.016$   & $20.5$   & $40*f_{orb}$ \tabularnewline
$f_{62}$  & & $388.995$   & $0.015$   & $2570.730$   & $0.099$   & $0.01280$   & $0.00071$   & $0.529$   & $0.018$   & $17.9$   & $41*f_{orb}$ \tabularnewline
$f_{63}$  & & $398.493$   & $0.015$   & $2509.456$   & $0.097$   & $0.01223$   & $0.00070$   & $0.904$   & $0.018$   & $17.4$   & $42*f_{orb}$ \tabularnewline
$f_{80}$  & & $407.955$   & $0.022$   & $2451.25$   & $0.13$   & $0.00855$   & $0.00070$   & $0.318$   & $0.026$   & $12.3$   & $43*f_{orb}$ \tabularnewline
$f_{90}$  & & $417.452$   & $0.030$   & $2395.49$   & $0.17$   & $0.00632$   & $0.00070$   & $0.687$   & $0.035$   & $9.1$   & $44*f_{orb}$ \tabularnewline
$f_{104}$  & & $426.923$   & $0.050$   & $2342.34$   & $0.27$   & $0.00373$   & $0.00069$   & $0.058$   & $0.059$   & $5.4$   & $45*f_{orb}$ \tabularnewline
$f_{98}$  & & $455.457$   & $0.036$   & $2195.60$   & $0.17$   & $0.00496$   & $0.00066$   & $0.628$   & $0.043$   & $7.5$   & $48*f_{orb}$ \tabularnewline
$f_{91}$  & & $464.888$   & $0.029$   & $2151.06$   & $0.14$   & $0.00604$   & $0.00066$   & $0.076$   & $0.035$   & $9.1$   & $49*f_{orb}$ \tabularnewline
$f_{75}$  & & $474.402$   & $0.018$   & $2107.917$   & $0.080$   & $0.00965$   & $0.00065$   & $0.432$   & $0.021$   & $14.8$   & $50*f_{orb}$ \tabularnewline
$f_{78}$  & & $483.886$   & $0.019$   & $2066.602$   & $0.082$   & $0.00908$   & $0.00065$   & $0.804$   & $0.023$   & $14.0$   & $51*f_{orb}$ \tabularnewline
$f_{79}$  & & $493.369$   & $0.019$   & $2026.882$   & $0.079$   & $0.00902$   & $0.00065$   & $0.200$   & $0.023$   & $13.9$   & $52*f_{orb}$ \tabularnewline
$f_{82}$  & & $502.862$   & $0.021$   & $1988.617$   & $0.085$   & $0.00814$   & $0.00065$   & $0.594$   & $0.025$   & $12.5$   & $53*f_{orb}$ \tabularnewline
$f_{84}$  & & $512.378$   & $0.022$   & $1951.686$   & $0.084$   & $0.00782$   & $0.00064$   & $0.928$   & $0.026$   & $12.2$   & $54*f_{orb}$ \tabularnewline
$f_{92}$  & & $521.822$   & $0.030$   & $1916.36$   & $0.11$   & $0.00578$   & $0.00064$   & $0.364$   & $0.035$   & $9.0$   & $55*f_{orb}$ \tabularnewline
$f_{106}$  & & $531.331$   & $0.049$   & $1882.06$   & $0.17$   & $0.00354$   & $0.00064$   & $0.680$   & $0.058$   & $5.5$  &$56*f_{orb}$ \tabularnewline
$f_{101}$  & & $569.322$   & $0.042$   & $1756.48$   & $0.13$   & $0.00412$   & $0.00064$   & $0.654$   & $0.050$   & $6.4$   & $60*f_{orb}$ \tabularnewline
$f_{102}$  & & $578.831$   & $0.044$   & $1727.62$   & $0.13$   & $0.00395$   & $0.00065$   & $0.020$   & $0.052$   & $6.1$   & $61*f_{orb}$ \tabularnewline
$f_{93}$  & & $588.300$   & $0.030$   & $1699.814$   & $0.088$   & $0.00576$   & $0.00065$   & $0.398$   & $0.036$   & $8.9$   & $62*f_{orb}$ \tabularnewline
$f_{95}$  & & $597.764$   & $0.031$   & $1672.902$   & $0.088$   & $0.00553$   & $0.00065$   & $0.798$   & $0.037$   & $8.5$   & $63*f_{orb}$ \tabularnewline
$f_{94}$  & & $607.202$   & $0.031$   & $1646.900$   & $0.083$   & $0.00563$   & $0.00065$   & $0.288$   & $0.036$   & $8.7$ & $64*f_{orb}$ \tabularnewline
$f_{100}$  & & $616.746$   & $0.036$   & $1621.412$   & $0.095$   & $0.00480$   & $0.00065$   & $0.582$   & $0.043$   & $7.4$   & $65*f_{orb}$ \tabularnewline
$f_{103}$  & & $626.243$   & $0.045$   & $1596.82$   & $0.12$   & $0.00383$   & $0.00065$   & $0.977$   & $0.054$   & $5.9$   & $66*f_{orb}$ \tabularnewline
$f_{105}$  & & $702.075$   & $0.048$   & $1424.349$   & $0.097$   & $0.00366$   & $0.00065$   & $0.533$   & $0.057$   & $5.6$   & $74*f_{orb}$ \tabularnewline
$f_{107}$  & & $721.098$   & $0.050$   & $1386.774$   & $0.097$   & $0.00346$   & $0.00065$   & $0.218$   & $0.060$   & $5.3$ & $76*f_{orb}$ \tabularnewline
$f_{108}$  & & $730.570$   & $0.051$   & $1368.793$   & $0.095$   & $0.00342$   & $0.00065$   & $0.648$   & $0.060$   & $5.3$  & $77*f_{orb}$ \tabularnewline
\tabularnewline
\multicolumn{3}{l}{Stellar pulsations} \tabularnewline
$f_{10}$  & & $22.026$   & $0.010$   & $45401$   & $21$   & $0.0668$   & $0.0025$   & $0.864$   & $0.012$   & $26.3$   & $F_2$ \tabularnewline
$f_{49}$  & & $23.454$   & $0.042$   & $42636$   & $76$   & $0.0161$   & $0.0025$   & $0.818$   & $0.050$   & $6.4$   &  \tabularnewline
$f_{35}$  & & $24.315$   & $0.031$   & $41127$   & $52$   & $0.0220$   & $0.0025$   & $0.615$   & $0.036$   & $8.8$   &  \tabularnewline
$f_{60}$  & & $28.030$   & $0.050$   & $35677$   & $64$   & $0.0132$   & $0.0025$   & $0.727$   & $0.060$   & $5.3$   &  \tabularnewline
$f_{29}$  & & $29.023$   & $0.027$   & $34455$   & $32$   & $0.0246$   & $0.0025$   & $0.512$   & $0.032$   & $10.0$   & $F_5$ ?  \tabularnewline
$f_{61}$  & & $35.837$   & $0.049$   & $27904$   & $38$   & $0.0129$   & $0.0024$   & $0.408$   & $0.058$   & $5.5$   & \tabularnewline
$f_{55}$  & & $43.978$   & $0.044$   & $22738$   & $23$   & $0.0138$   & $0.0023$   & $0.139$   & $0.052$   & $6.1$   &  \tabularnewline
$f_{23}$  & & $87.144$   & $0.016$   & $11475.3$   & $2.2$   & $0.0301$   & $0.0018$   & $0.277$   & $0.019$   & $16.4$   & $F_7$ \tabularnewline
$f_{38}$  & & $89.865$   & $0.023$   & $11127.8$   & $2.9$   & $0.0210$   & $0.0018$   & $0.117$   & $0.028$   & $11.5$   &  \tabularnewline
$f_{50}$  & & $95.848$   & $0.030$   & $10433.1$   & $3.3$   & $0.0159$   & $0.0018$   & $0.489$   & $0.036$   & $8.9$   &  \tabularnewline
$f_{57}$  & & $96.753$   & $0.035$   & $10335.6$   & $3.8$   & $0.0137$   & $0.0018$   & $0.516$   & $0.042$   & $7.6$   &  \tabularnewline
$f_{58}$  & & $100.122$   & $0.036$   & $9987.8$   & $3.6$   & $0.0135$   & $0.0018$   & $0.860$   & $0.042$   & $7.5$   &  \tabularnewline
$f_{52}$  & & $106.280$   & $0.032$   & $9409.1$   & $2.8$   & $0.0148$   & $0.0018$   & $0.339$   & $0.038$   & $8.4$   &  \tabularnewline
$f_{54}$  & & $109.238$   & $0.034$   & $9154.3$   & $2.8$   & $0.0140$   & $0.0018$   & $0.025$   & $0.040$   & $7.9$   & \tabularnewline
$f_{70}$  & & $114.628$   & $0.041$   & $8723.8$   & $3.1$   & $0.0112$   & $0.0017$   & $0.922$   & $0.049$   & $6.5$   &  \tabularnewline
$f_{72}$  & & $118.771$   & $0.043$   & $8419.6$   & $3.1$   & $0.0106$   & $0.0017$   & $0.575$   & $0.051$   & $6.2$   &  \tabularnewline
$f_{30}$  & & $124.654$   & $0.019$   & $8022.2$   & $1.2$   & $0.0241$   & $0.0017$   & $0.466$   & $0.022$   & $14.4$   & $F_{12}$ \tabularnewline
$f_{64}$  & & $125.360$   & $0.037$   & $7977.0$   & $2.3$   & $0.0122$   & $0.0017$   & $0.788$   & $0.044$   & $7.3$   &  \tabularnewline
$f_{77}$  & & $135.183$   & $0.047$   & $7397.4$   & $2.6$   & $0.0093$   & $0.0016$   & $0.334$   & $0.056$   & $5.7$   &  \tabularnewline
$f_{67}$  & & $136.861$   & $0.037$   & $7306.7$   & $2.0$   & $0.0116$   & $0.0016$   & $0.999$   & $0.044$   & $7.2$   & $F_{13}$ \tabularnewline
$f_{87}$  & & $210.808$   & $0.045$   & $4743.7$   & $1.0$   & $0.0069$   & $0.0012$   & $0.941$   & $0.053$   & $6.0$   & $F_{23}$ \tabularnewline
$f_{66}$  & & $222.306$   & $0.025$   & $4498.31$   & $0.51$   & $0.0116$   & $0.0011$   & $0.102$   & $0.030$   & $10.6$   & $F_{15}$ \tabularnewline
$f_{74}$  & & $231.121$   & $0.028$   & $4326.73$   & $0.52$   & $0.0102$   & $0.0011$   & $0.945$   & $0.033$   & $9.7$   & $F_{17}$ \tabularnewline
$f_{9}$  & & $233.3617$   & $0.0036$   & $4285.193$   & $0.067$   & $0.0775$   & $0.0010$   & $0.6100$   & $0.0043$   & $74.1$   & $F_1$ \tabularnewline
$f_{20}$  & & $237.7339$   & $0.0083$   & $4206.38$   & $0.15$   & $0.0329$   & $0.0010$   & $0.7959$   & $0.0098$   & $32.4$   & $F_6$ \tabularnewline
$f_{46}$  & & $242.707$   & $0.016$   & $4120.19$   & $0.27$   & $0.0171$   & $0.0010$   & $0.257$   & $0.019$   & $17.1$   & $F_{11}$ \tabularnewline
$f_{13}$  & & $243.7941$   & $0.0047$   & $4101.822$   & $0.079$   & $0.05673$   & $0.00100$   & $0.1163$   & $0.0056$   & $56.9$   & $F_4$ \tabularnewline
$f_{81}$  & & $251.541$   & $0.031$   & $3975.50$   & $0.49$   & $0.00836$   & $0.00096$   & $0.650$   & $0.037$   & $8.7$   & $F_{19}$ \tabularnewline
$f_{96}$  & & $253.147$   & $0.047$   & $3950.28$   & $0.74$   & $0.00542$   & $0.00096$   & $0.804$   & $0.056$   & $5.7$   & \tabularnewline
$f_{99}$  & & $262.651$   & $0.051$   & $3807.33$   & $0.74$   & $0.00487$   & $0.00092$   & $0.362$   & $0.060$   & $5.3$   & $F_{27}$ \tabularnewline
$f_{69}$  & & $268.599$   & $0.021$   & $3723.02$   & $0.29$   & $0.01135$   & $0.00090$   & $0.871$   & $0.025$   & $12.7$   & $F_{16}$ \tabularnewline
$f_{28}$  & & $288.0106$   & $0.0091$   & $3472.09$   & $0.11$   & $0.02478$   & $0.00084$   & $0.523$   & $0.011$   & $29.4$   & $F_9$ \tabularnewline
$f_{68}$  & & $296.724$   & $0.019$   & $3370.13$   & $0.22$   & $0.01140$   & $0.00082$   & $0.012$   & $0.023$   & $13.9$   & $F_{14}$ \tabularnewline
$f_{76}$  & & $317.796$   & $0.023$   & $3146.67$   & $0.22$   & $0.00939$   & $0.00079$   & $0.961$   & $0.027$   & $11.9$   & $F_{18}$ \tabularnewline
$f_{27}$  & & $321.6966$   & $0.0083$   & $3108.519$   & $0.080$   & $0.02519$   & $0.00078$   & $0.1757$   & $0.0098$   & $32.4$   & $F_8$ \tabularnewline
$f_{97}$  & & $348.785$   & $0.040$   & $2867.10$   & $0.33$   & $0.00503$   & $0.00075$   & $0.505$   & $0.047$   & $6.8$   & $F_{26}$ \tabularnewline
$f_{32}$  & & $394.9676$   & $0.0085$   & $2531.853$   & $0.055$   & $0.02234$   & $0.00071$   & $0.974$   & $0.010$   & $31.6$   & $F_{10}$ \tabularnewline
\label{listfreq}
\end{longtable}
\normalsize

\FloatBarrier
\pagebreak

\section{Transit fit correlations}
\begin{figure*}[h]
\centering
\includegraphics[width=\linewidth]{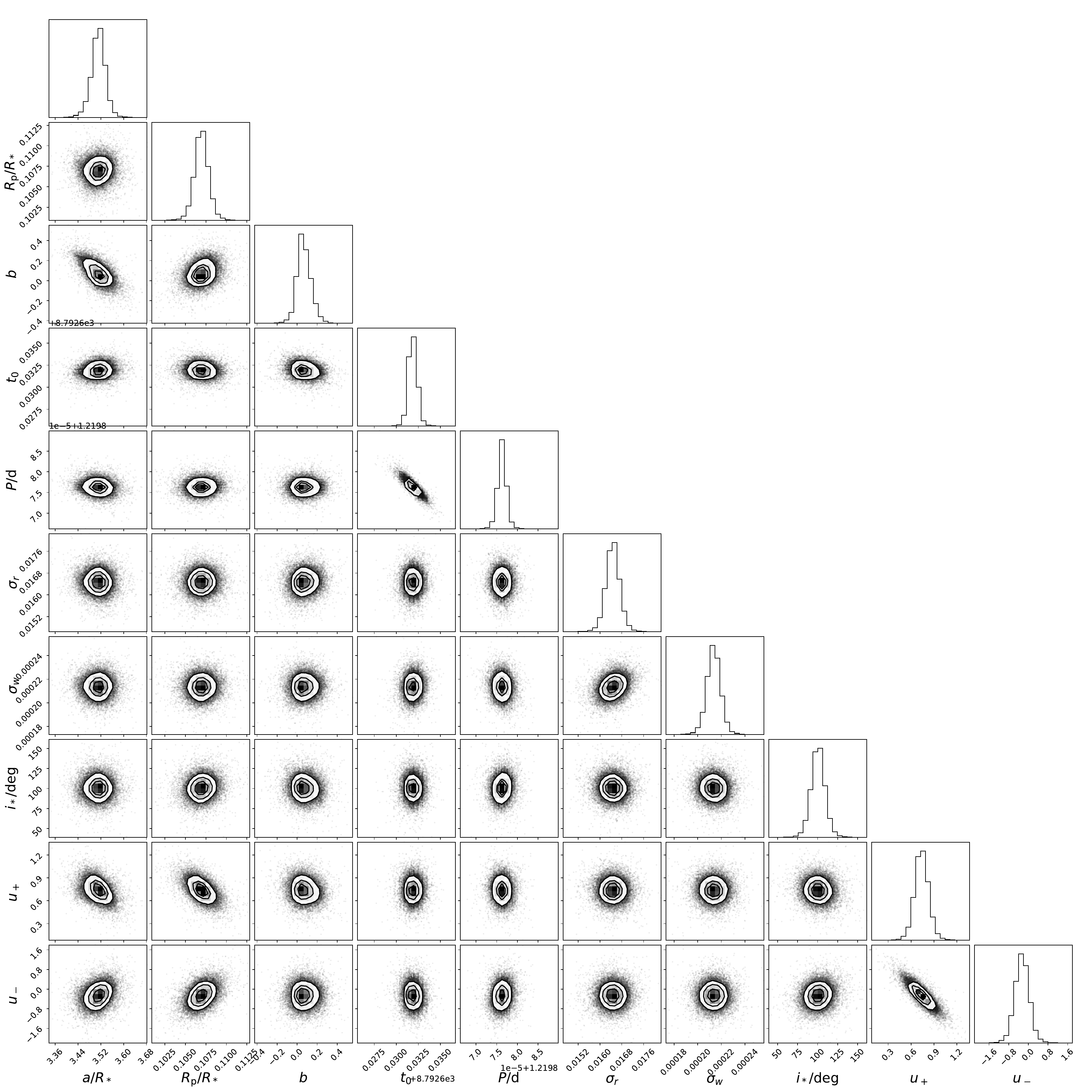}
\caption{Posterior distribution from the case (iii) transit fit (See Section ~\ref{sec:transits}).}
\label{fig:corner} 
\end{figure*} 
\FloatBarrier
\pagebreak

\section{Example occultation fits}
\begin{figure*}[h]
\centering
\includegraphics[angle=270,width=0.5\linewidth]{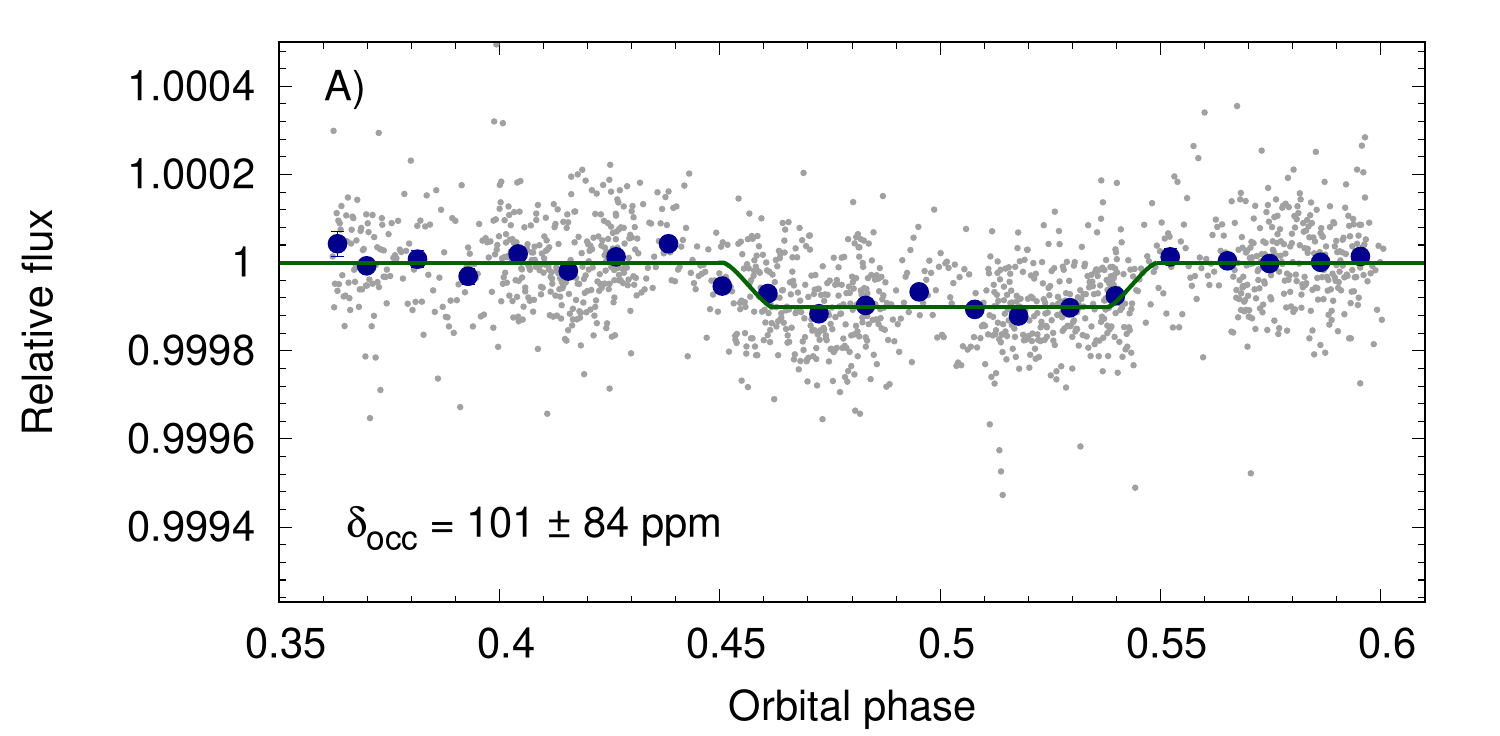}
\includegraphics[angle=270,width=0.5\linewidth]{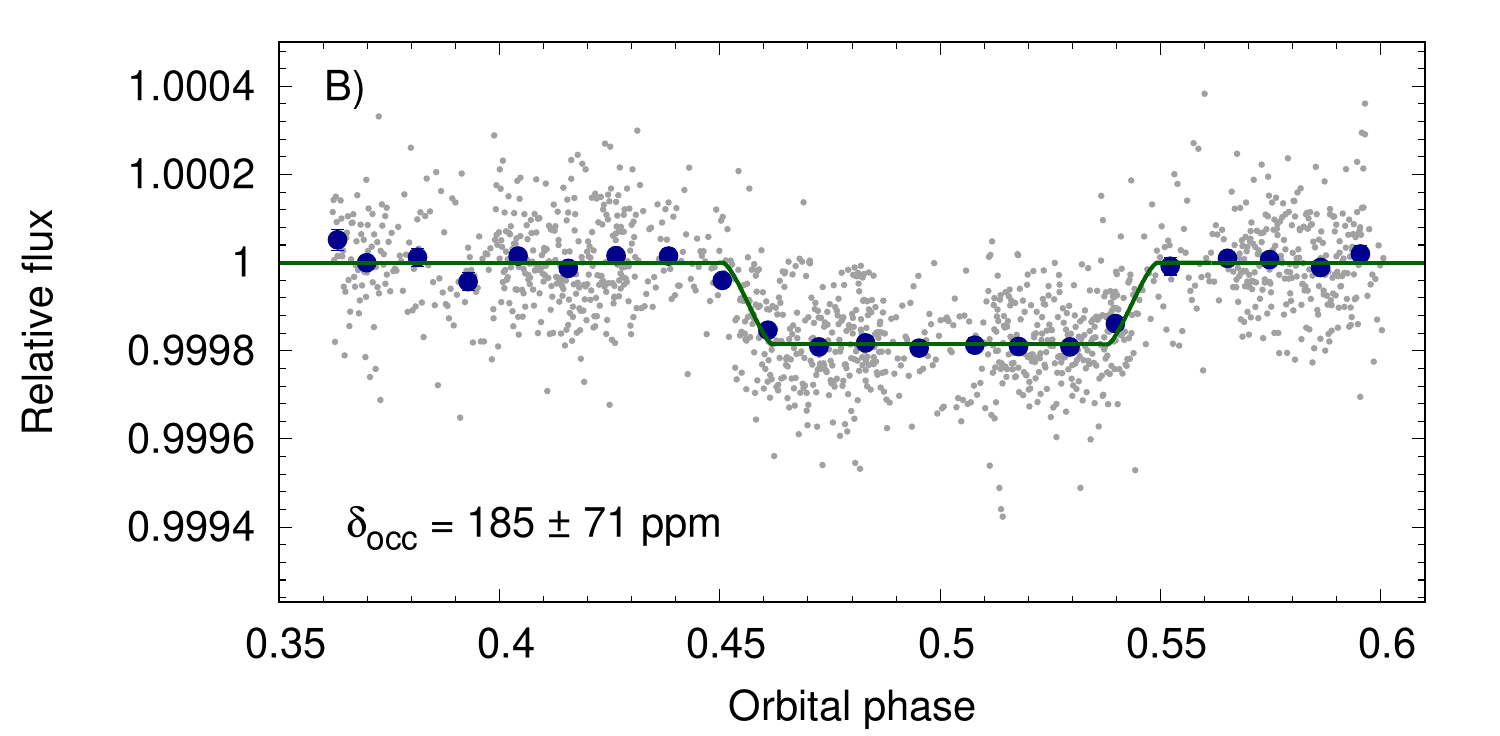}
\includegraphics[angle=270,width=0.5\linewidth]{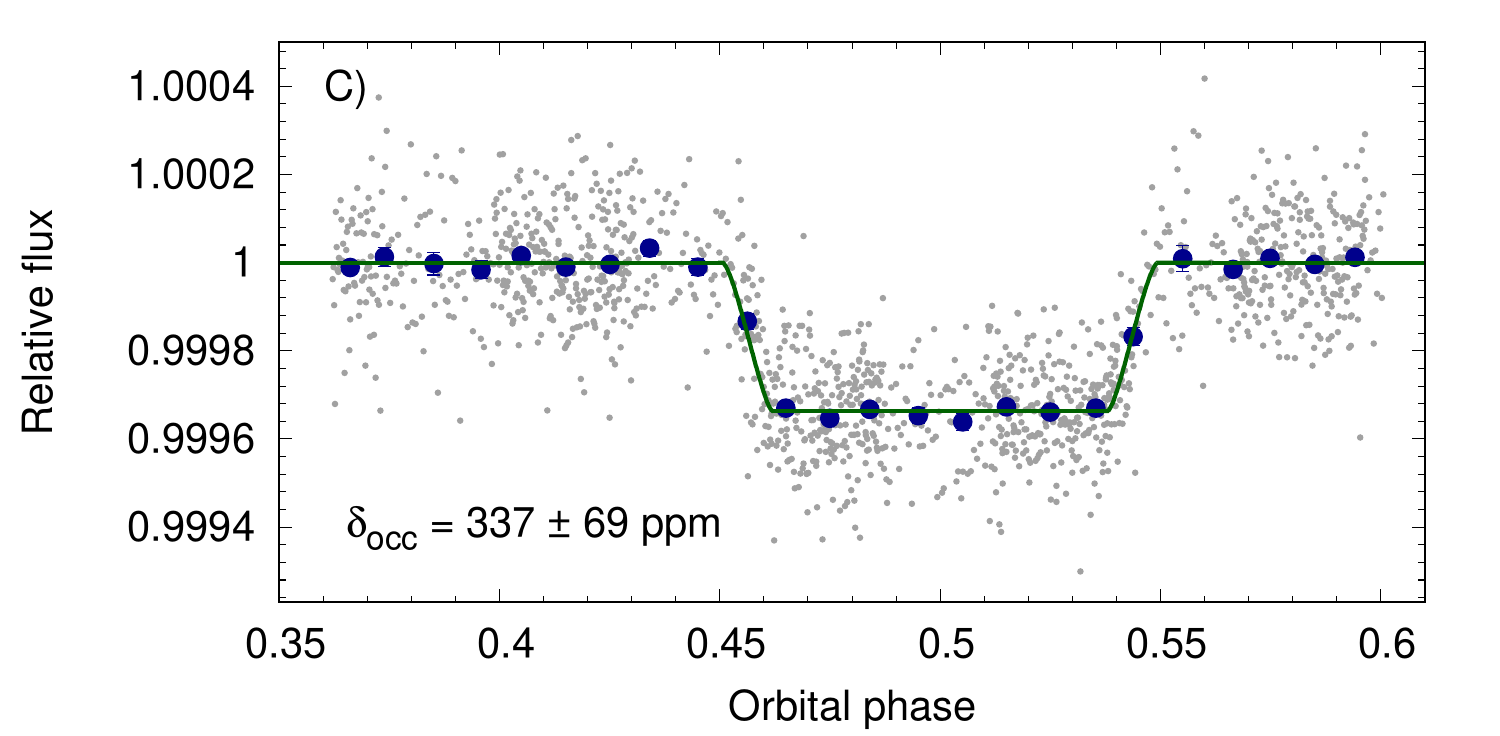}
\includegraphics[angle=270,width=0.5\linewidth]{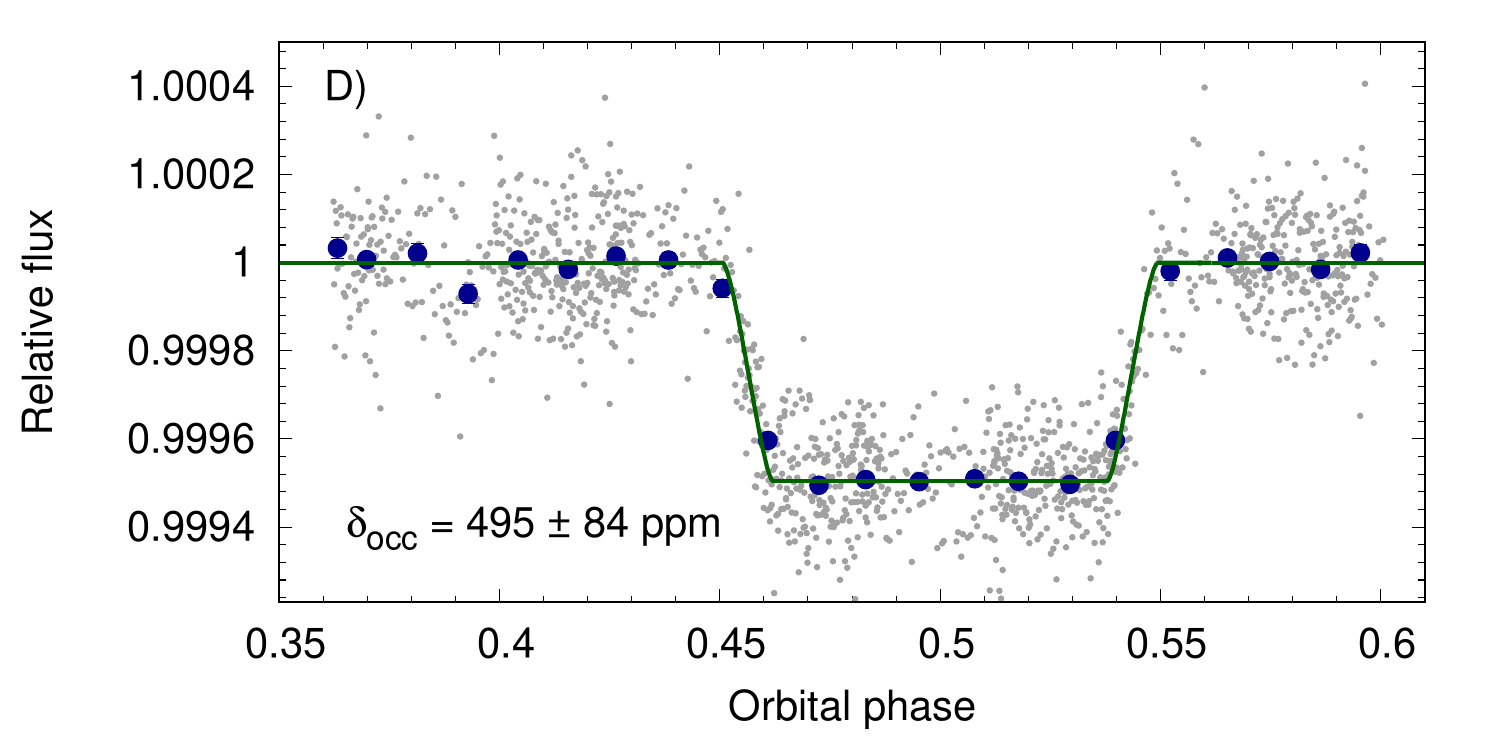}
\caption{Examples of some of our fits to the CHEOPS occultation data, showing a range of occultation depths. In each panel, the grey points represent the unbinned data with the pulsation signal subtracted, the larger, dark blue points are binned to 0.01 in orbital phase (about 17.6 minutes), and the green line is our best-fitting model for this particular treatment of the pulsations. A) three frequencies (from \citep{w33_Kalman_2022}) per occultation. B) 10 frequencies (from \citep{w33_vonEssen_2020}) fitted in common to all four occultations. C) as (B), but for 20 frequencies. D) 15 frequencies from this work (Section~\ref{sec:puls_tess}).}
\label{fig:occ_apdx} 
\end{figure*} 
\FloatBarrier
\pagebreak

\section{Silhouette plots for clustering analysis}
\label{sec:silhouette}

\begin{figure*}[h]
\centering
\includegraphics[angle=0,width=0.9\linewidth]{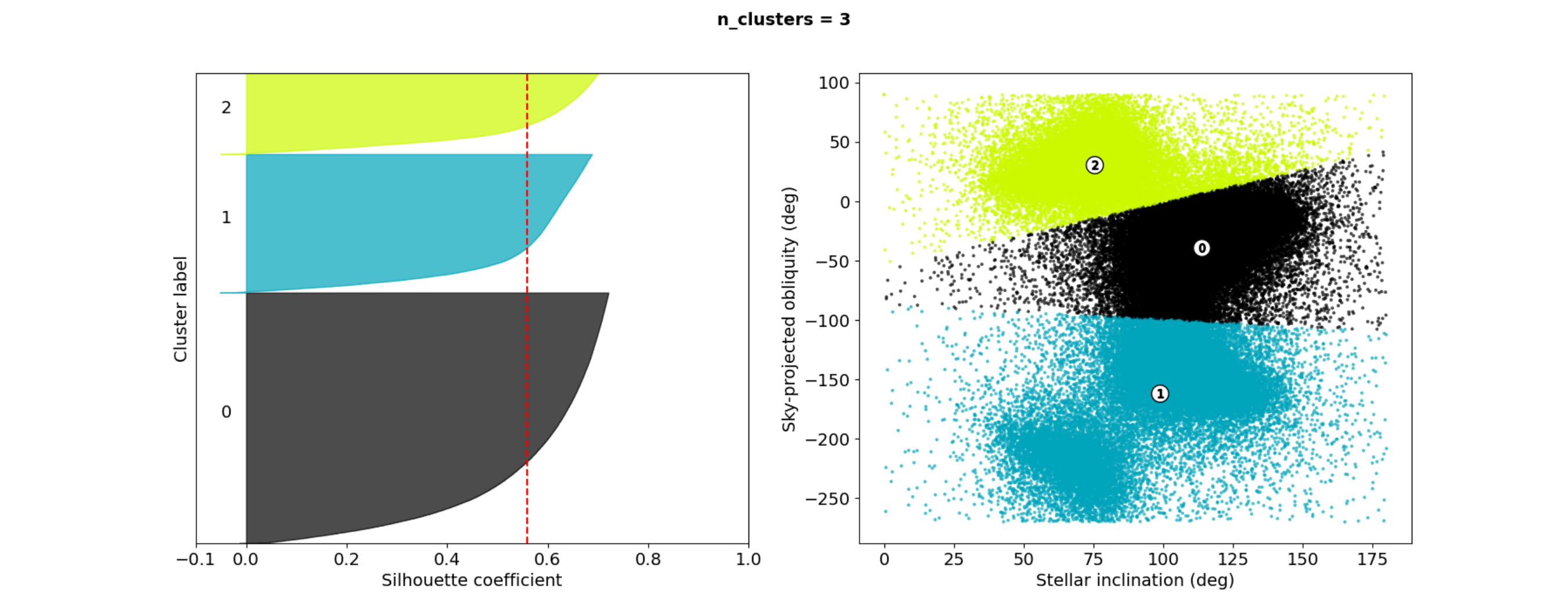}
\includegraphics[angle=0,width=0.9\linewidth]{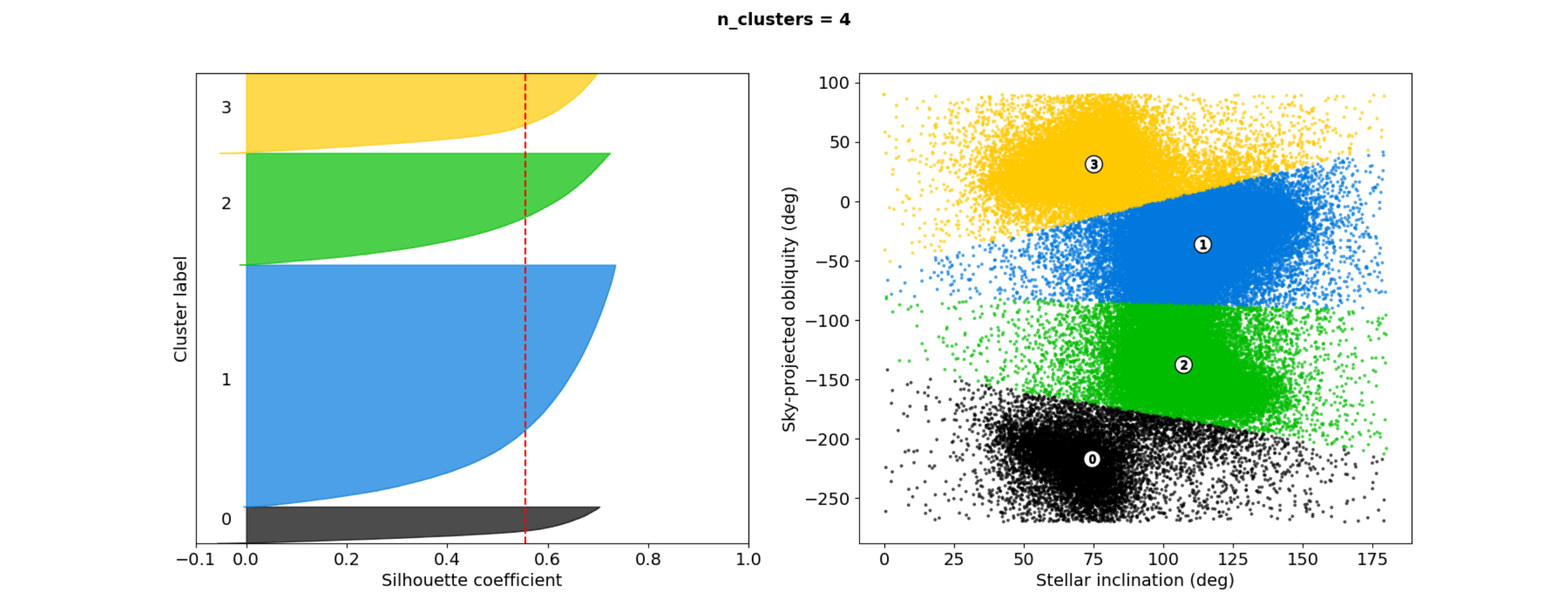}
\includegraphics[angle=0,width=0.9\linewidth]{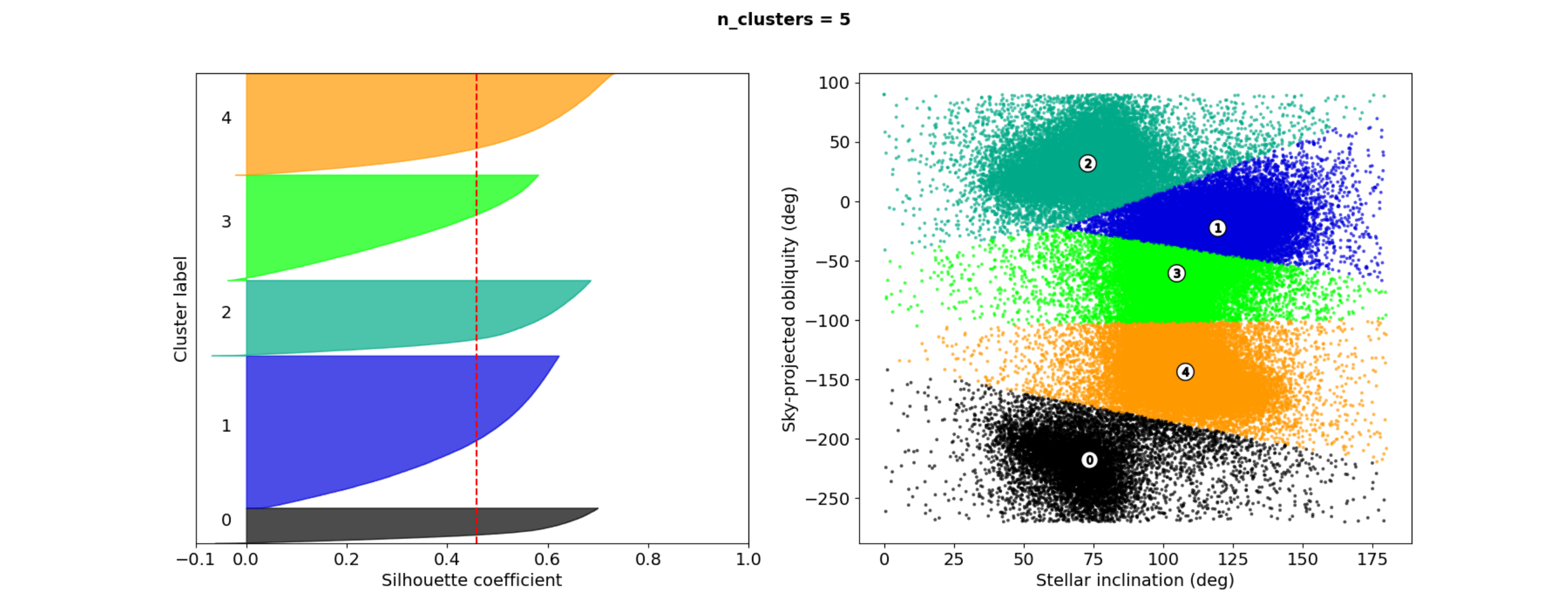}
\caption{Silhouette plots for our cluster analysis of the posterior distribution of \Omstar and \istar. Silhouette plots are shown (from top to bottom) for three, four, and five clusters. In each case, the silhouette coefficients are plotted in the left panel, where the silhouette statistic is indicated with a dashed red line. The clusters are shown in $\lambda$ -- \istar space in the right panels, using the same colour coding as in the corresponding left-hand panel. For more details, see Section~\ref{sec:results_transit}.}
\label{fig:silhouette} 
\end{figure*} 

\end{appendix}
\end{document}